\documentclass[12pt]{article}

\usepackage[margin=1in]{geometry}
\usepackage{amsmath, amssymb, amsthm}
\usepackage{graphicx}
\usepackage{natbib}
\usepackage{booktabs}
\usepackage{algorithm}
\usepackage{algpseudocode}
\usepackage[hidelinks]{hyperref}
\usepackage{xcolor}
\usepackage{enumitem}
\usepackage{bm}
\usepackage{array}
\newcolumntype{R}[1]{>{\raggedleft\arraybackslash}p{#1}}
\newcolumntype{C}[1]{>{\centering\arraybackslash}p{#1}}
\usepackage{tikz}
\usetikzlibrary{shapes.geometric, arrows.meta, positioning, fit, backgrounds, calc}

\newtheorem{theorem}{Theorem}
\newtheorem{proposition}{Proposition}

\newtheorem{corollary}{Corollary}

\newtheorem{assumption}{}

\newcommand{\E}{\mathbb{E}}
\newcommand{\R}{\mathbb{R}}
\newcommand{\Prob}{\mathbb{P}}
\newcommand{\Var}{\mathrm{Var}}

\newcommand{\X}{\mathcal{X}}

\newcommand{\calT}{\mathcal{T}}
\newcommand{\calL}{\mathcal{L}}

\newcommand{\wh}[1]{\widehat{#1}}
\newcommand{\wt}[1]{\widetilde{#1}}
\newcommand{\argmin}{\operatornamewithlimits{argmin}}
\newcommand{\argmax}{\operatornamewithlimits{argmax}}

\title{\textbf{Co-Supervised Tree Synthesis for Interpretable Subgroup Identification and Honest Treatment Effect Inference in Randomized Trials}}

\author{
Evgenii Kuriabov$^1$, \quad Zhaoxue Tong$^2$, \quad Jia Li$^1$ \\[6pt]
{\normalsize $^1$Department of Statistics, The Pennsylvania State University} \\
{\normalsize $^2$Department of Statistics, Florida State University} \\[4pt]
{\normalsize Correspondence: Evgenii Kuriabov, \href{mailto:eak5582@psu.edu}{eak5582@psu.edu}. ORCID: \href{https://orcid.org/0009-0007-2879-6791}{0009-0007-2879-6791}}
}

\date{}

\begin{document}
\maketitle

\begin{abstract}
Identifying patient subgroups with heterogeneous treatment effects is central to precision medicine, yet existing approaches face a tension. Black-box methods estimate individualized effects well but yield no interpretable subgroups, while tree-based methods produce explicit partitions but are unstable and less accurate in moderate-sample trials.
We propose CausalSynthTree, a co-supervised method in which black-box causal estimators guide construction of an interpretable tree.
It partitions the covariate space into cells, fits cell-wise conditional average treatment effect models from both observed data and synthetic data labeled by black-box causal teachers, and synthesizes them into a tree through a treatment-effect disparity criterion.
Each leaf carries a sparse linear model, so the tree path defines a subgroup and the leaf model shows which covariates modify the effect.
Honest inference for leaf-wise effects remains valid despite co-supervised augmentation and data-driven subgroup selection.
In simulations it closes much of the gap between the two families. From moderate sample sizes onward it is more accurate than the black-box learners whenever the effect is linear or has a single threshold, and trails them only on a finer partition than the tree it grows. It controls spurious splits and attains nominal coverage throughout.
In the ACTG175 HIV trial, where standard interaction tests detect no effect modification, the method reports no subgroups while the competing tree methods report several. In an adjuvant colon cancer trial it reports a borderline covariate in part of the resamples, together with a within-region age gradient that is consistent with an independent test.

\end{abstract}

\section{Introduction}
\label{sec:intro}

A recurring question in clinical trials is whether the overall treatment effect masks meaningful heterogeneity. Some patients may benefit substantially while others derive little or no benefit, or may even be harmed.
Identifying such subgroups has implications for personalized treatment recommendations, trial design, regulatory labeling, and resource allocation.
The ICH E9(R1) addendum on estimands \citep{ICHE9R1} and the FDA guidance on enrichment strategies \citep{FDA2019} both emphasize the importance of understanding treatment effect heterogeneity, yet practical tools that balance statistical rigor with clinical interpretability remain limited.

Existing methods for subgroup identification vary along two axes, the degree of model flexibility, from parametric to nonparametric, and the form of the output, an interpretable partition or pointwise estimates.
At one extreme, interaction trees \citep{Su2009}, SIDES \citep{Lipkovich2011}, and model-based recursive partitioning \citep[MOB;][]{Zeileis2008, Seibold2016} build tree partitions directly from the observed data, yielding clinically communicable subgroup rules that can be unstable in moderate-sample settings and do not leverage the predictive power of modern machine learning.
At the other extreme, causal forests \citep{Wager2018} and related ensemble methods provide flexible, consistent estimation of the \emph{conditional average treatment effect} (CATE) \citep{Kuenzel2019, Nie2021} with valid pointwise inference, but they produce no interpretable partition, and clinicians cannot easily describe ``who benefits'' from a kernel-weighted average over thousands of trees.

Between these extremes, the modified covariate method \citep{Tian2014} offers a clean reparameterization that isolates treatment-covariate interactions, but it produces a single global linear model rather than a partition.
Change-plane methods \citep{Fan2017} identify a single linear boundary separating responders from non-responders, which may be too restrictive for complex heterogeneity patterns.
None of these methods simultaneously delivers an interpretable partition, within-region effect modification, and valid leaf-wise inference, which is the gap CausalSynthTree fills.

This paper proposes CausalSynthTree, a co-supervised method for subgroup identification.
Black-box causal estimators (causal forests, T-learners, or X-learners \citep{Kuenzel2019}) serve as teachers that label synthetic data with treatment effects, and these labels are combined with unbiased pseudo-outcomes from the trial to fit local CATE models that remain stable in cells with limited observed data.
The idea of co-supervising local models with a black-box teacher originates in SynthTree \citep{Kuriabov2024}, a method for explainable prediction.
The causal setting shapes every stage.
Since the treatment effect is never observed, unbiased pseudo-outcomes stand in for the missing labels, adaptive weights decide how far each local model trusts the teacher, cells are aligned with the teacher's effect surface, the tree is grown on disagreement in treatment effects, and honest estimation yields valid confidence intervals for the subgroups it finds.

\subsection{Contributions}

The output of CausalSynthTree is a shallow tree whose leaves carry a sparse linear model for the treatment effect, so a practitioner reads off an explicit rule that defines a subgroup and a short formula showing which covariates modify the effect within it, with no need for separate post-hoc tools.
Three design choices make this work.
CATE-weighted $k$-means \citep{Lloyd1982} aligns cell boundaries with the teacher's effect surface, adaptive per-cell weights recalibrate reliance on the teacher by its disagreement with unbiased pseudo-outcomes and protect the cell models where it is unreliable, and a final refinement step moves each split threshold from the cell centroids onto the observed data.

The output carries guarantees.
Honest leaf-wise confidence intervals attain their nominal level despite the synthetic augmentation used to build the tree, because the entire co-supervision pipeline is confined to the construction sample.
A closed-form bias-variance decomposition says exactly when co-supervision lowers cell-level error, namely when the teacher's local bias is smaller than the variance it removes, and the adaptive weights are a tractable proxy for the optimal trade-off.
The tree-synthesis criterion is consistent for the true partition and shares the split-quality structure of generalized random forests \citep{Athey2019GRF}.

Empirically it closes much of the gap between the two families.
From a thousand observations onward it is more accurate than the black-box learners whenever a sharp threshold separates two regions, with or without a linear drift within them, and it trails them only when the true partition is finer than the tree it grows.
When the effect is linear with no subgroups it returns a single leaf, and the sparse linear model in that leaf estimates the effect more accurately than any of the black-box learners.
It is also the most parsimonious partition method at every sample size, collapsing to one or two leaves under the null where competing trees report spurious subgroups.
Its weaknesses are stated with equal care.
When the true partition has several regions with similar effects the method merges some of them and a larger tree recovers the finer partition better, and detection at clinical noise levels requires large subgroup contrasts.

\subsection{Related Work}
\label{sec:related}

We briefly position CausalSynthTree relative to the main families of subgroup identification methods, for which \citet{Lipkovich2017} and \citet{Lipkovich2024} give comprehensive taxonomies.

Tree-based methods build partitions directly from observation-level criteria.
Interaction trees replace the impurity of CART \citep{Breiman1984} with an interaction statistic, SIDES searches for regions of enhanced effect, MOB tests for parameter instability in a pre-specified model, and causal trees \citep{Athey2016} target the MSE of the CATE with honest estimation.
CausalSynthTree instead evaluates candidate splits on pre-estimated cell-wise CATE models.
Ensemble and black-box methods, among them causal forests, BART \citep{Chipman2010}, and metalearners, provide flexible CATE estimation with no interpretable partition, and any of them can serve as a co-supervising teacher.
Model-based methods such as pairwise fusion \citep{Ma2017, Zhang2019} and structured mixtures \citep{Shen2015} model heterogeneity directly but do not produce tree-structured partitions, which are often preferred for clinical communication.
For inference after selection, \citet{GuoHe2020} develop de-biased bootstrap procedures for the best selected subgroup, which we apply to honest leaf estimates.

The closest relatives are distillation methods.
Virtual Twins \citep{Foster2011} fits a CART tree to estimates of individual effects from a random forest \citep{Breiman2001}, a two-stage approach in which the second tree is built on noisy estimates and the CART criterion does not directly target effect heterogeneity.
Causal distillation trees \citep[CDT;][]{Huang2025} likewise fit CART to teacher-predicted effects but add split consistency guarantees and a stability-based teacher selection criterion.
CausalSynthTree differs from both in three respects.
It produces sparse linear leaf models capturing within-region effect modification rather than constant subgroup effects.
It uses co-supervised local estimation with adaptive weights rather than pure distillation, anchoring estimates in observed pseudo-outcomes and down-weighting the teacher under disagreement.
And it constructs cells by CATE-weighted clustering before tree synthesis and splits on cell-wise CATE models rather than pointwise estimates, inheriting the consistent split structure of generalized random forests.
SynthTree and the MLM-EPIC framework \citep{Seo2022} build explainable surrogates for prediction under black-box co-supervision.
CausalSynthTree brings co-supervision to treatment effects, where the quantity being learned is never observed.
The teacher is therefore calibrated against unbiased pseudo-outcomes rather than against outcomes, the tree is grown on disagreement in treatment effects, and the honest split gives valid inference for the subgroups the tree selects.

The remainder of this article is organized as follows.
Section~\ref{sec:setup} formalizes the subgroup identification problem.
Section~\ref{sec:method} presents the five-stage CausalSynthTree procedure and its inference tools.
Section~\ref{sec:theory} states the theoretical guarantees, whose proofs appear in the Supporting Information.
Section~\ref{sec:simulations} reports the simulation studies, Section~\ref{sec:realdata} presents the two trial applications, and Section~\ref{sec:discussion} concludes.
The Supporting Information is available online.

\section{Problem Setup}
\label{sec:setup}

Consider a randomized clinical trial with $n$ subjects.
For each subject $i$, let $X_i \in \R^p$ denote the vector of baseline covariates, $T_i \in \{0, 1\}$ the binary treatment assignment (0 = control, 1 = treatment), and $Y_i \in \R$ the observed outcome.
Under the potential outcomes framework \citep{Rubin1974, Neyman1923}, each subject has two potential outcomes: $Y_i(1)$ under treatment and $Y_i(0)$ under control, with the observed outcome $Y_i = T_i Y_i(1) + (1 - T_i) Y_i(0)$.

\begin{assumption}[Randomization and overlap]
\label{ass:randomization}
Treatment is randomized: $T_i \perp (Y_i(1), Y_i(0)) \mid X_i$, and the propensity score $e(x) = \Prob(T = 1 \mid X = x)$ satisfies $0 < \eta \le e(x) \le 1 - \eta < 1$ for some $\eta > 0$.
In a randomized trial with equal allocation, $e(x) = 1/2$.
\end{assumption}

The quantity of primary interest is the CATE:
\begin{equation}
\label{eq:cate}
    \tau(x) = \E[Y(1) - Y(0) \mid X = x].
\end{equation}
A subgroup is a region $S \subseteq \X$ defined by interpretable rules on $X$, within which the average treatment effect $\E[\tau(X) \mid X \in S]$ differs meaningfully from the population average $\E[\tau(X)]$.

We formalize subgroup identification as the problem of finding a partition $\mathcal{P} = \{P_1, \ldots, P_K\}$ of the covariate space $\X$ such that within each region $P_k$, the CATE is well-approximated by a parsimonious model:
\begin{equation}
\label{eq:mlm_cate}
    \tau(x) \approx \wt\tau(x) = \sum_{k=1}^{K} \mathbf{1}(x \in P_k) \, \tau_k(x), \qquad \tau_k(x) = \alpha_k + x^\top \beta_k.
\end{equation}
When $\beta_k = 0$ for all $k$, this reduces to a piecewise-constant CATE model (as in causal trees).
By allowing sparse $\beta_k \ne 0$, CausalSynthTree captures within-region effect modification while keeping the model interpretable.

The partition $\mathcal{P}$ is further constrained to be tree-structured. Each $P_k$ corresponds to a leaf of a binary decision tree $\calT$ with axis-aligned splits $X_j \le c$ in the original feature space, and we write $\calL(\calT) = \{P_1, \ldots, P_K\}$ for its set of leaves.
This ensures that subgroup membership can be communicated as a sequence of simple threshold rules.

\section{Methodology}
\label{sec:method}

\begin{figure}[t]
\centering
\resizebox{\textwidth}{!}{%
\begin{tikzpicture}[
    >=Stealth,
    node distance=0.5cm and 0.8cm,
    databox/.style={rectangle, draw=black!70, fill=blue!8, rounded corners=3pt,
        minimum height=0.7cm, minimum width=2.4cm, align=center, font=\small},
    procbox/.style={rectangle, draw=black!70, fill=orange!10, rounded corners=3pt,
        minimum height=0.7cm, minimum width=2.4cm, align=center, font=\small},
    stagebox/.style={rectangle, draw=black!80, fill=yellow!12, rounded corners=3pt,
        minimum height=0.8cm, minimum width=2.8cm, align=center, font=\small\bfseries},
    resultbox/.style={rectangle, draw=black!80, fill=green!10, rounded corners=3pt,
        minimum height=0.7cm, minimum width=2.4cm, align=center, font=\small},
    treebox/.style={rectangle, draw=black!80, fill=green!15, rounded corners=5pt,
        minimum height=1.0cm, minimum width=4.5cm, align=center, font=\small},
    arr/.style={->, thick, black!70},
    darr/.style={->, thick, dashed, black!50},
    stagelabel/.style={font=\footnotesize\sffamily\bfseries, text=black!60},
]

\node[databox, minimum width=3.8cm] (input) {Observed data $(X, T, Y)$};

\node[procbox, below left=0.7cm and 0.3cm of input] (teachers) {Train CATE teachers\\[-2pt]{\footnotesize (CF, T/S/X-learner)}};
\node[procbox, below right=0.7cm and 0.3cm of input] (cluster) {CATE-weighted $k$-means\\[-2pt]{\footnotesize select $(J, \rho)$ by $R$-loss}};

\draw[arr] (input.south) -- ++(0,-0.25) -| (teachers.north);
\draw[arr] (input.south) -- ++(0,-0.25) -| (cluster.north);

\node[stagelabel, left=0.1cm of teachers] {Stage 1};
\node[stagelabel, right=0.1cm of cluster] {Stage 2};

\node[databox, below=0.5cm of teachers] (teacher_out) {$\widehat{\tau}^{\mathrm{BB}}(x)$};
\node[databox, below=0.5cm of cluster] (cells_out) {Cells $C_1, \ldots, C_J$};

\draw[arr] (teachers) -- (teacher_out);
\draw[arr] (cluster) -- (cells_out);

\node[stagebox, below=1.15cm of $(teacher_out)!0.5!(cells_out)$, minimum width=5.2cm] (stage3) {
    Co-supervised local CATE estimation\\[-2pt]
    {\footnotesize Pseudo-outcomes $Z_i^{\mathrm{aug}}$ $+$ synthetic data $\to$ Lasso per cell}
};
\node[stagelabel, left=0.1cm of stage3] {Stage 3};

\draw[arr] (teacher_out.south) -- ++(0,-0.3) -| ([xshift=-0.5cm]stage3.north);
\draw[arr] (cells_out.south) -- ++(0,-0.3) -| ([xshift=0.5cm]stage3.north);

\node[databox, below=0.5cm of stage3, minimum width=4.5cm] (cell_models) {Cell-wise CATE models $g_1(x), \ldots, g_J(x)$};
\draw[arr] (stage3) -- (cell_models);

\node[stagebox, below=0.5cm of cell_models, minimum width=5.2cm] (stage4) {
    Tree synthesis via CATE disparity\\[-2pt]
    {\footnotesize Greedy splits maximizing $\Delta(t, f, c)$}
};
\node[stagelabel, left=0.1cm of stage4] {Stage 4};
\draw[arr] (cell_models) -- (stage4);

\node[stagebox, below=0.5cm of stage4, minimum width=5.2cm] (stage5) {
    Leaf refitting, sizing \& refinement\\[-2pt]
    {\footnotesize Refit sparse CATE model per leaf; CC-prune; refine thresholds}
};
\node[stagelabel, left=0.1cm of stage5] {Stage 5};
\draw[arr] (stage4) -- (stage5);

\node[treebox, below=0.6cm of stage5] (output) {
    \textbf{CausalSynthTree}\\[-1pt]
    {\footnotesize Tree paths $=$ subgroup rules \quad $|$ \quad Leaf models $=$ sparse $\tau_k(x)$}
};
\draw[arr] (stage5) -- (output);

\node[resultbox, right=0.6cm of output, minimum width=2.5cm] (inference) {Honest inference\\[-2pt]{\footnotesize (estimation sample)}};
\draw[darr] (output.east) -- (inference.west);

\end{tikzpicture}%
}
\caption{Schematic illustration of the CausalSynthTree pipeline.
Observed data feed into two parallel streams: black-box CATE teacher training (Stage~1) and covariate clustering (Stage~2).
These are combined in Stage~3 to produce co-supervised cell-wise CATE models, which are then synthesized into a tree via a disparity criterion (Stage~4) and refined by leaf refitting, pruning, and threshold refinement (Stage~5).
The final tree provides interpretable subgroup definitions through its paths and sparse within-region CATE models at its leaves.
Honest inference is performed on a held-out estimation sample.}
\label{fig:schematic}
\end{figure}
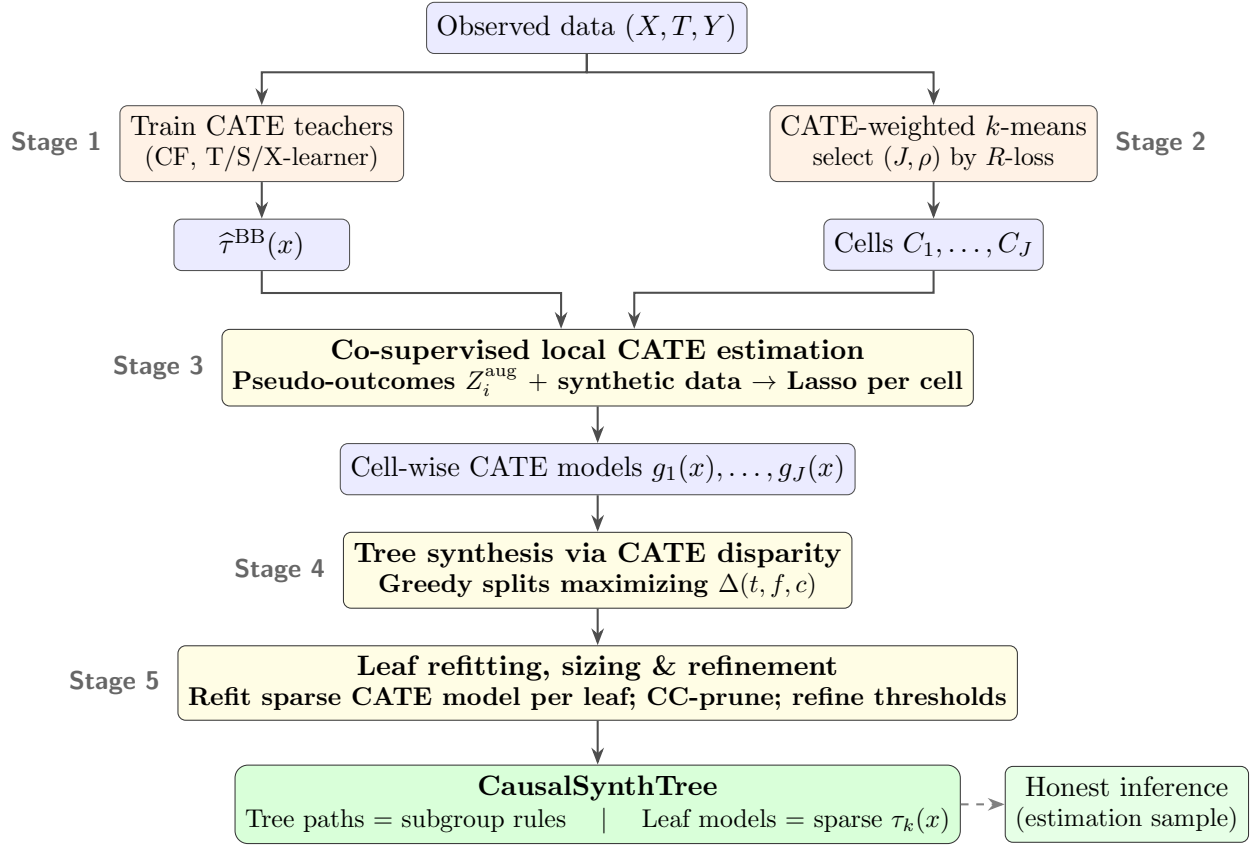

The CausalSynthTree algorithm proceeds in five stages, summarized in Algorithm~\ref{alg:causal_synthtree} and illustrated schematically in Figure~\ref{fig:schematic}.

\subsection{Stage 1: Black-box CATE Teachers}
\label{sec:teachers}

We train one or more black-box CATE estimators on the observed data $\{(x_i, t_i, y_i)\}_{i=1}^n$.
Any estimator that produces a function $\wh\tau^{\mathrm{BB}}: \R^p \to \R$ can serve as a teacher.
The implementation supports the following teachers:
\begin{itemize}[nosep]
    \item \textbf{T-learner:} fit separate response surfaces $\wh\mu_1(x)$ and $\wh\mu_0(x)$ for treated and control groups, then set $\wh\tau^{\mathrm{T}}(x) = \wh\mu_1(x) - \wh\mu_0(x)$.
    \item \textbf{S-learner} \citep{Kuenzel2019}: fit a single model $\wh\mu(x, t)$ including treatment as a feature, then set $\wh\tau^{\mathrm{S}}(x) = \wh\mu(x, 1) - \wh\mu(x, 0)$.
    \item \textbf{X-learner:} a two-stage cross-imputation approach that is particularly effective when treatment and control groups have different sizes.
    \item \textbf{Causal forest:} an ensemble of honest causal trees with pointwise CATE estimation and valid confidence intervals.
\end{itemize}
Several teachers may be combined. All experiments in this paper use a single teacher, a causal forest except where a T-learner is compared with it in Section~\ref{sec:sim_sensitivity}, and Section~\ref{sec:discussion} discusses multi-teacher extensions.

\subsection{Stage 2: CATE-weighted Clustering}
\label{sec:clustering}

Standard $k$-means clustering on the covariate vectors $\{x_i\}$ produces cells that are spatially compact but may straddle treatment effect boundaries. A cell near a CATE discontinuity can contain observations from both sides, blurring the treatment effect signal and weakening the downstream disparity criterion.
To address this, we propose \emph{CATE-weighted $k$-means}, which clusters on an augmented feature space that incorporates the teacher's CATE predictions.

Define the augmented feature vector for each observation:
\begin{equation}
\label{eq:cate_augment}
    \tilde{x}_i^{(\rho)} = \left(\frac{x_i - \bar{x}}{\wh\sigma_x},\; \rho \cdot \frac{\wh\tau^{\mathrm{BB}}(x_i) - \overline{\wh\tau}}{\wh\sigma_{\wh\tau}}\right) \in \R^{p+1},
\end{equation}
where the covariate and CATE components are standardized to have zero mean and unit variance, and $\rho \ge 0$ controls the relative importance of CATE alignment versus spatial coherence.
When $\rho = 0$, this reduces to standard $k$-means on covariates. When $\rho > 0$, cells are encouraged to be CATE-homogeneous, aligning cell boundaries with the teacher's treatment effect surface.

We apply $k$-means to $\{\tilde{x}_i^{(\rho)}\}_{i=1}^n$ to obtain $J$ cells $C_1, \ldots, C_J$.
The cell centroids in the original covariate space (the first $p$ components) are used for tree splitting in Stage~4.
The pair $(J, \rho)$ is selected jointly by validation. We split the training data into sub-training and validation sets, construct the initial piecewise CATE model (Section~\ref{sec:local_cate}) for each candidate pair, and select the $(J, \rho)$ that minimizes the validation-set $R$-loss.

To evaluate CATE prediction error without observing individual treatment effects, we use the $R$-loss of \citet{Nie2021}:
\begin{equation}
\label{eq:rloss}
    \wh{\calL}_R(\wt\tau) = \frac{1}{n_{\mathrm{val}}} \sum_{i \in \mathrm{val}} \left\{ Y_i - \wh\mu(X_i) - (T_i - \wh{e}(X_i)) \, \wt\tau(X_i) \right\}^2,
\end{equation}
where $\wh\mu(x)$ is an estimate of $\E[Y \mid X = x]$ and $\wh{e}(x)$ is an estimate of the propensity score.
In a randomized trial the propensity score is known by design, so only $\wh\mu$ is estimated, and we use the same pilot estimate that enters the pseudo-outcomes of Section~\ref{sec:local_cate}.

The inclusion of $\rho = 0$ in the candidate set provides a built-in safeguard. If the teacher's CATE predictions are uninformative or misleading, the validation procedure selects $\rho = 0$ and recovers standard covariate-based clustering.

\subsection{Stage 3: Co-supervised Local CATE Estimation}
\label{sec:local_cate}

This co-supervised local estimation is the core of the method. Within each cell $C_j$, we estimate a local linear CATE model $g_j(x) = \alpha_j + x^\top \beta_j$ using both observed data and synthetic data generated by the black-box teacher.
Concretely, the training set for $g_j$ is the union of (i) the observed points falling in $C_j$, paired with pseudo-outcomes, and (ii) the synthetic points drawn from $C_j$'s empirical distribution, labeled by the teacher.

\paragraph{Observed-data component.}
In the spirit of the modified outcome approach, we construct pseudo-outcomes that isolate the treatment effect.
With a known propensity score $e(x)$, define
\begin{equation}
\label{eq:pseudo_outcome}
    Z_i = \frac{T_i - e(X_i)}{e(X_i)\,\{1 - e(X_i)\}} \, Y_i,
\end{equation}
which satisfies $\E[Z_i \mid X_i] = \tau(X_i)$ exactly, because $\E[(T - e(X))\,Y \mid X] = e(X)\{1 - e(X)\}\,\tau(X)$ under \ref{ass:randomization} (Lemma~B.1 in Appendix~B of the Supporting Information).
For a randomized trial with $e(x) = 1/2$, this reduces to $Z_i = 2\,(2T_i - 1)\, Y_i$.

An efficiency-augmented version \citep{Liang2020} subtracts a pilot estimate of the conditional mean outcome:
\begin{equation}
\label{eq:pseudo_augmented}
    Z_i^{\mathrm{aug}} = \frac{T_i - e(X_i)}{e(X_i)\,\{1 - e(X_i)\}} \big(Y_i - \wh\mu(X_i)\big),
\end{equation}
where $\wh\mu(x)$ is a pilot estimate of $\mu(x) = \E[Y \mid X = x]$ (e.g., from a random forest on pooled data).
The subtracted term enters through the multiplier $(T_i - e(X_i))/[e(X_i)\{1 - e(X_i)\}]$, whose conditional mean given $X_i$ is zero, so $\E[Z_i^{\mathrm{aug}} \mid X_i] = \tau(X_i)$ for any $\wh\mu$ that does not depend on $(T_i, Y_i)$.
The pilot is therefore cross-fitted \citep{Chernozhukov2018}, and $\wh\mu(X_i)$ is the out-of-fold prediction of a forest trained with the fold containing observation $i$ held out.
An in-sample pilot would absorb part of $(T_i - e(X_i))\,\tau(X_i)$ into $\wh\mu(X_i)$ and attenuate $Z_i^{\mathrm{aug}}$ toward zero, which shrinks every downstream leaf model.
The augmentation therefore preserves unbiasedness while reducing variance when $\wh\mu$ is close to $\mu$.
Lemma~B.1 in Appendix~B of the Supporting Information gives the conditional variance of $Z_i^{\mathrm{aug}}$, which is minimized at $\wh\mu = \mu$.

\paragraph{Co-supervised augmentation.}
For each cell $C_j$ with centroid $\bar{x}_j$ and within-cell sample covariance diagonal $\mathrm{diag}(s_{j,1}^2, \ldots, s_{j,p}^2)$, we generate $n_{\mathrm{syn}}$ synthetic covariate vectors:
\begin{equation}
\label{eq:synth}
    x^{\mathrm{syn}} \sim \mathcal{N}\big(\bar{x}_j, \, \mathrm{diag}(s_{j,1}^2, \ldots, s_{j,p}^2)\big).
\end{equation}
Each synthetic point is labeled with the teacher's CATE prediction: $z^{\mathrm{syn}} = \wh\tau^{\mathrm{BB}}(x^{\mathrm{syn}})$.
This augmentation serves two purposes. It stabilizes local estimation when the cell contains few observed data points, and it transfers the black-box model's knowledge of local treatment effect structure into the cell-wise linear model.

\paragraph{Adaptive co-supervision weights.}
The two signals feeding $g_j$ sit at opposite ends of the bias-variance spectrum. The pseudo-outcome $Z_i^{\mathrm{aug}}$ is unbiased for $\tau(X_i)$ but has high per-observation variance, whereas the teacher $\wh\tau^{\mathrm{BB}}$ is low-variance because it smooths across the full sample but can be biased in any given cell. A fixed weight cannot accommodate this, since teacher quality varies across cells and the teacher cannot audit its own bias. We therefore measure teacher reliability against the only unbiased signal available. Conditional on the teacher, $\E[(Z_i^{\mathrm{aug}} - \wh\tau^{\mathrm{BB}}(X_i))^2 \mid X_i] = \Var(Z_i^{\mathrm{aug}} \mid X_i) + (\tau(X_i) - \wh\tau^{\mathrm{BB}}(X_i))^2$, so the average of this squared disagreement within $C_j$ estimates the pseudo-outcome variance plus the teacher's local squared bias, and the pseudo-outcomes act as a noisy but unbiased yardstick for the teacher.
Define the local discrepancy
\begin{equation}
\label{eq:local_discrepancy}
    D_j = \frac{1}{|C_j|} \sum_{i \in C_j} \big(Z_i^{\mathrm{aug}} - \wh\tau^{\mathrm{BB}}(x_i)\big)^2,
\end{equation}
which measures how much the teacher's predictions deviate from the pseudo-outcomes in cell $C_j$, and the local pseudo-outcome variance $V_j = \Var(Z_i^{\mathrm{aug}} \mid i \in C_j)$.
The adaptive weight for cell $C_j$ is
\begin{equation}
\label{eq:adaptive_weight}
    w_{\mathrm{syn},j} = \mathrm{clip}\!\left(\frac{V_j}{V_j + D_j},\; 0.1,\; 0.9\right), \qquad w_{\mathrm{obs},j} = 1 - w_{\mathrm{syn},j}.
\end{equation}
When $D_j$ is small (teacher agrees with data), $w_{\mathrm{syn},j}$ is large and the cell model leans on the teacher for variance reduction.
When $D_j$ is large (teacher disagrees), $w_{\mathrm{syn},j}$ decreases, protecting the cell model from teacher bias.
Corollary~\ref{cor:helps_main} in Section~\ref{sec:theory} identifies the bias-variance-optimal weight, and the ratio in \eqref{eq:adaptive_weight} is a plug-in proxy for it.
Because $D_j$ contains the irreducible variance $V_j$ as well as the squared bias, the ratio caps the weight of even a perfect teacher near one half.
A bias-corrected discrepancy would replace $D_j$ by $D_j - \wh V_j$, where $\wh V_j$ is a within-cell estimate of the pseudo-outcome noise variance, and would let an accurate teacher receive a weight close to one while a biased teacher is still down-weighted.
The clipped ratio is the default used throughout, and Appendix~F of the Supporting Information shows that it matches or beats both the bias-corrected rule and a plug-in oracle weight, because the cap near one half guards against importing the teacher's attenuated within-region slopes into the leaf models.

\paragraph{Local CATE model fitting.}
We combine the observed pseudo-outcomes $\{(x_i, Z_i^{\mathrm{aug}})\}_{i \in C_j}$ and the synthetic teacher-labeled data $\{(x_k^{\mathrm{syn}}, z_k^{\mathrm{syn}})\}_{k=1}^{n_{\mathrm{syn}}}$, and fit a Lasso regression \citep{Tibshirani1996}:
\begin{multline}
\label{eq:local_lasso}
    (\wh\alpha_j, \wh\beta_j) = \argmin_{\alpha, \beta} \bigg\{
        \frac{w_{\mathrm{obs},j}}{|C_j|} \sum_{i \in C_j} (Z_i^{\mathrm{aug}} - \alpha - x_i^\top \beta)^2 \\
        + \frac{w_{\mathrm{syn},j}}{n_{\mathrm{syn}}} \sum_{k=1}^{n_{\mathrm{syn}}} (z_k^{\mathrm{syn}} - \alpha - (x_k^{\mathrm{syn}})^\top \beta)^2
        + \lambda_j \|\beta\|_1
    \bigg\},
\end{multline}
where $\lambda_j$ is the Lasso penalty selected by cross-validation within the cell.

The collection of cell-wise models yields an initial piecewise CATE estimate:
\begin{equation}
\label{eq:initial_mlm}
    \wt\tau(x) = \sum_{j=1}^{J} \mathbf{1}(x \in C_j)\, g_j(x).
\end{equation}

\subsection{Stage 4: Tree Synthesis via CATE Disparity}
\label{sec:tree}

The tree synthesis stage grows a tree over the cell-wise CATE models by greedy splitting on a disparity criterion.
The greedy synthesis of local models into a tree is shared with SynthTree, and the criterion, which compares treatment-effect predictions rather than outcome predictions, is what makes the resulting partition a subgroup partition.

\paragraph{CATE disparity.}
The disparity between two cell-wise CATE models $g_j$ and $g_{j'}$ is defined as the mean squared difference in their treatment effect predictions over the pooled support of the two cells:
\begin{equation}
\label{eq:cate_disparity}
    d(g_j, g_{j'}) = \frac{1}{|S_j \cup S_{j'}|} \sum_{x \in S_j \cup S_{j'}} \big(g_j(x) - g_{j'}(x)\big)^2,
\end{equation}
where $S_j$ is the set of original and augmented data points used to estimate $g_j$.
This measures how similarly two cell-wise CATE models describe treatment effect heterogeneity on their combined support.
Models that assign similar treatment effects to the same patients should be merged. Models that disagree should be separated.

\paragraph{Node impurity and goodness of split.}
For a node $t$ containing cell-wise CATE models $g_j$, $j \in \mathcal{J}_t$, define the impurity:
\begin{equation}
\label{eq:impurity}
    D(t) = \frac{1}{\binom{|\mathcal{J}_t|}{2}} \sum_{j < j' \in \mathcal{J}_t} d(g_j, g_{j'}).
\end{equation}
A candidate split partitions the models into left and right child nodes $t_L$ and $t_R$ based on a threshold $X_f \le c$ applied to the cell centroids.
The goodness of split is:
\begin{equation}
\label{eq:gos}
    \Delta(t, f, c) = D(t) - p_L \cdot D(t_L) - p_R \cdot D(t_R),
\end{equation}
where $p_L = |\mathcal{J}_{t_L}| / |\mathcal{J}_t|$ and $p_R = |\mathcal{J}_{t_R}| / |\mathcal{J}_t|$.
The split $(f^*, c^*)$ maximizing $\Delta$ is selected at each node.
Splitting stops when no candidate yields $\Delta > 0$ or when a maximum depth $d_{\max}$ is reached.

\paragraph{Interpretation of the disparity criterion.}
Unlike outcome-level impurity (CART) or parameter instability tests (MOB), the CATE disparity criterion directly measures disagreement in \emph{treatment effect predictions} among pre-estimated local models.
This suits subgroup identification for two reasons. It separates cells where treatment effects differ even when the patients are prognostically similar, and it is invariant to shifts in the baseline outcome level, so prognostic heterogeneity alone does not inflate the disparity.
The criterion has the same structure as the $\Xi$-criterion of generalized random forests, which measures heterogeneity by the squared difference in parameter estimates between candidate child nodes and is optimal there in the sense of minimizing downstream estimation error.
The difference is that CausalSynthTree evaluates it on pre-estimated cell-wise models rather than on individual observations.

\subsection{Stage 5: Leaf Refitting, Tree Sizing, and Threshold Refinement}
\label{sec:refit}

Once the full tree is grown, we refit a sparse CATE model in each leaf.
Rather than assigning data to leaves based on cell-centroid membership, we route every data point, both observed and synthetic, through the tree's split rules to determine its leaf assignment.
This ensures consistency between the tree structure and the leaf models.

Specifically, we first generate synthetic covariate vectors from all $J$ cells (as in Stage~3) and label them with the teacher's CATE predictions.
We then route all observed points $\{x_i\}_{i=1}^n$ and all synthetic points through the tree to determine their leaf assignments.
For each leaf $P_k$, let $P_k^{\mathrm{obs}} = \{i : x_i \text{ routes to } P_k\}$ and $P_k^{\mathrm{syn}} = \{l : x_l^{\mathrm{syn}} \text{ routes to } P_k\}$.
We fit a co-supervised Lasso:
\begin{multline}
\label{eq:leaf_refit}
    (\wh\alpha_k, \wh\beta_k) = \argmin_{\alpha, \beta} \bigg\{
        \frac{w_{\mathrm{obs},k}}{|P_k^{\mathrm{obs}}|} \sum_{i \in P_k^{\mathrm{obs}}} (Z_i^{\mathrm{aug}} - \alpha - x_i^\top \beta)^2 \\
        + \frac{w_{\mathrm{syn},k}}{|P_k^{\mathrm{syn}}|} \sum_{l \in P_k^{\mathrm{syn}}} (z_l^{\mathrm{syn}} - \alpha - (x_l^{\mathrm{syn}})^\top \beta)^2
        + \lambda_k \|\beta\|_1
    \bigg\}.
\end{multline}

\paragraph{Tree sizing.}
We determine the final tree size by cost-complexity pruning \citep{Breiman1984}.
Starting from the fully grown tree, we generate a nested sequence of subtrees by iteratively collapsing the internal node whose removal least increases the $R$-loss.
The optimal subtree is selected by $V$-fold cross-validated $R$-loss \eqref{eq:rloss} on the construction data.
Details of the pruning procedure are provided in Appendix~C of the Supporting Information.

\paragraph{Threshold refinement.}
Every split of the pruned tree sits at the midpoint between two cell centroids, so its location is resolved only to the spacing of the cells.
As a final step we move each surviving threshold to the observed data.
Consider a split on $X_f$, and let $c_L$ be the largest centroid coordinate on $X_f$ among the cells in the left child and $c_R$ the smallest among the cells in the right child, so that $c_L < c_R$.
The candidates are the midpoints between consecutive observed values of $X_f$ in $(c_L, c_R)$ among the construction points routed to the node, and we select the candidate that minimizes the pseudo-outcome sum of squares under the two child models.
Thresholds are refined from the root down and the leaf models are then refit by \eqref{eq:leaf_refit}.
Refinement follows pruning, so whether a split exists is decided on the coarser centroid tree, and refinement only sharpens the location of the splits that survive.


\begin{algorithm}[t]
\caption{CausalSynthTree Training Procedure}
\label{alg:causal_synthtree}
\begin{algorithmic}[1]
\Require Training data $(X, T, Y)$; candidate cell counts $\mathcal{J}$; candidate CATE-weighting parameters $\mathcal{R}$; black-box CATE teacher(s); maximum depth $d_{\max}$.
\Ensure A fitted CausalSynthTree with leaf-wise sparse CATE models.
\Statex
\State \textbf{Stage 1:} Train black-box CATE teacher(s) on $(X, T, Y)$; obtain $\wh\tau^{\mathrm{BB}}(x)$.
\State Compute the cross-fitted pilot estimate $\wh\mu(x)$ for the augmented pseudo-outcomes \eqref{eq:pseudo_augmented}.
\Statex
\For{each candidate $(J, \rho) \in \mathcal{J} \times \mathcal{R}$}
    \State \textbf{Stage 2:} Form augmented features $\tilde{x}_i^{(\rho)}$ via \eqref{eq:cate_augment}; cluster into $J$ cells via $k$-means.
    \For{each cell $C_j$}
        \State \textbf{Stage 3a:} Compute pseudo-outcomes $Z_i^{\mathrm{aug}}$ and adaptive weights $(w_{\mathrm{obs},j}, w_{\mathrm{syn},j})$ via \eqref{eq:adaptive_weight}.
        \State \textbf{Stage 3b:} Generate synthetic covariates $x^{\mathrm{syn}} \sim \mathcal{N}(\bar{x}_j, \mathrm{diag}(s_j^2))$; label with teacher CATE.
        \State \textbf{Stage 3c:} Fit Lasso CATE model $g_j(x)$ on combined data with adaptive weights.
    \EndFor
    \State Form initial piecewise CATE model $\wt\tau(x)$; evaluate on validation set via $R$-loss.
\EndFor
\State Select $(J^*, \rho^*)$ with best validation $R$-loss; rebuild cells and models on full training data.
\Statex
\State \textbf{Stage 4:} Grow tree on cell-wise CATE models using disparity-based goodness of split \eqref{eq:gos}.
\State Stop splitting when $\Delta \le 0$ or maximum depth $d_{\max}$ reached.
\Statex
\State \textbf{Stage 5:} Route all observed and synthetic points through the tree; refit sparse CATE model in each leaf via \eqref{eq:leaf_refit}.
\State Select final tree size by CC-prune using cross-validated $R$-loss.
\State Refine each surviving split threshold on the observed data; refit the leaf models via \eqref{eq:leaf_refit}.
\end{algorithmic}
\end{algorithm}

\subsection{Honest Estimation for Leaf-wise Treatment Effects}
\label{sec:honest}

A key requirement for subgroup analysis in clinical trials is valid inference for treatment effects within the identified subgroups.
We adopt the honest estimation strategy of causal trees. The data are split into a \emph{construction sample} $\mathcal{S}^{\mathrm{con}}$ used to build the tree (Stages 1 to 5 above) and an \emph{estimation sample} $\mathcal{S}^{\mathrm{est}}$ used for final inference.
Given the tree structure $\calT$ determined from $\mathcal{S}^{\mathrm{con}}$, the leaf-wise average treatment effect in leaf $P_k$ is estimated from $\mathcal{S}^{\mathrm{est}}$ by:
\begin{equation}
\label{eq:honest_ate}
    \wh\tau_k^{\mathrm{est}} = \frac{1}{n_{k,1}} \sum_{\substack{i \in \mathcal{S}^{\mathrm{est}} \\ x_i \in P_k, \, t_i = 1}} Y_i - \frac{1}{n_{k,0}} \sum_{\substack{i \in \mathcal{S}^{\mathrm{est}} \\ x_i \in P_k, \, t_i = 0}} Y_i,
\end{equation}
where $n_{k,1}$ and $n_{k,0}$ are the counts of treated and control subjects in leaf $P_k$ within $\mathcal{S}^{\mathrm{est}}$.
Because $\mathcal{S}^{\mathrm{est}}$ is independent of the tree construction, $\wh\tau_k^{\mathrm{est}}$ is unbiased for the leaf-wise ATE $\E[\tau(X) \mid X \in P_k]$, and a standard two-sample confidence interval is valid:
\begin{equation}
\label{eq:ci}
    \wh\tau_k^{\mathrm{est}} \pm z_{1-\alpha/2} \sqrt{\frac{\wh\sigma_{k,1}^2}{n_{k,1}} + \frac{\wh\sigma_{k,0}^2}{n_{k,0}}},
\end{equation}
where $\wh\sigma_{k,t}^2$ is the sample variance of outcomes in leaf $P_k$ among subjects with treatment $t$.

\paragraph{Co-supervision and honesty.}
A subtlety is that the black-box teachers in Stage 1 are trained on the full construction sample $\mathcal{S}^{\mathrm{con}}$, and the synthetic augmentation in Stage 3 uses teacher predictions on synthetic covariates.
Because the estimation sample $\mathcal{S}^{\mathrm{est}}$ plays no role in teacher training, cell construction, tree synthesis, pruning, or threshold refinement, all of which use only construction-sample points and pseudo-outcomes, the tree structure, including the synthetic augmentation that shaped it, is a deterministic function of $\mathcal{S}^{\mathrm{con}}$.
Inference in \eqref{eq:ci} then depends only on $\mathcal{S}^{\mathrm{est}}$ given the tree, so the honesty of \eqref{eq:honest_ate} is preserved.

\subsection{Post-Selection Inference for the Best Subgroup}
\label{sec:postselection}

When the goal is to identify the subgroup with the largest treatment effect and provide a valid confidence interval for its effect size, we face the winner's curse, meaning the maximum of estimated effects is biased upward for the maximum of true effects \citep{GuoHe2020}.
We apply the de-biased bootstrap procedure of Guo and He to the leaf-wise estimates $\wh\tau_1^{\mathrm{est}}, \ldots, \wh\tau_K^{\mathrm{est}}$, which produces an asymptotically sharp lower confidence bound for the largest leaf-wise ATE that accounts for the selection of the best leaf.
Proposition~A.2 in the Supporting Information verifies the conditions of their result in this setting and shows that the naive one-sided bound covers at only $(1-\alpha)^m$ when $m$ leaves are tied at the maximum.

\subsection{Within-Leaf CATE Model Inference}
\label{sec:leaf_inference}

Beyond the leaf-wise ATE, CausalSynthTree's leaf models $\tau_k(x) = \alpha_k + x^\top \beta_k$ provide information about within-region effect modification.
For inference on the coefficient $\beta_k$, we adopt a post-Lasso approach \citep{Belloni2013}. After identifying the active set $\wh{A}_k = \{\ell : \wh\beta_{k,\ell} \ne 0\}$ from the Lasso fit, we refit an OLS regression of $Z_i^{\mathrm{aug}}$ on $x_{i, \wh{A}_k}$ using only the estimation sample and compute standard OLS confidence intervals.

The post-Lasso intervals for $\beta_k$ are valid conditional on the selected active set $\wh{A}_k$ and do not account for the selection step itself, so a covariate the Lasso excludes is reported as exactly zero even when its true coefficient is small but nonzero.
For unconditional coverage we also implement the de-biased Lasso \citep{Javanmard2014, Zhang2014}, which replaces the post-Lasso refit by a one-step correction based on a nodewise-Lasso estimate of the inverse Gram matrix, at a modest cost in interval width.
Appendix~E of the Supporting Information evaluates both procedures.
The leaf-wise ATE of Section~\ref{sec:honest} remains the primary subgroup-level output.

\section{Theoretical Properties}
\label{sec:theory}

The co-supervised construction raises two questions.
Does building the tree with synthetic, teacher-labeled data invalidate downstream inference, and when does leaning on the teacher help rather than hurt local estimation?
We answer them with three results. Theorem~\ref{thm:honest_main} shows that honest leaf-wise inference is unaffected by the co-supervised construction, Theorem~\ref{thm:cosup_main} and Corollary~\ref{cor:helps_main} give the cell-level error of the co-supervised fit and the condition under which the teacher lowers it, and Proposition~\ref{prop:disparity_main} shows that the disparity criterion recovers the true partition.
Each result is stated under the conditions of Section~\ref{sec:assumptions_main}.
The regularity conditions, the explicit constants, and all proofs are given in Appendices~A and~B of the Supporting Information.

\subsection{Assumptions}
\label{sec:assumptions_main}

Beyond the randomization condition \ref{ass:randomization}, the analysis uses five technical conditions.

\begin{assumption}[Bounded outcomes and smoothness]
\label{ass:bounded_main}
Outcomes are bounded, $|Y(t)| \le M$ for $t \in \{0,1\}$, the covariate space $\X \subset \R^p$ is compact, and $\tau$, the conditional means $\mu_t(x) = \E[Y(t) \mid X = x]$, and the conditional variances of $Y(t)$ are Lipschitz on $\X$.
\end{assumption}

\begin{assumption}[Teacher consistency]
\label{ass:teacher_main}
$\sup_{x \in \X} |\wh\tau^{\mathrm{BB}}(x) - \tau(x)| = O_p(r_n)$ for some rate $r_n \to 0$.
\end{assumption}

\begin{assumption}[Local linearity on a tree-structured partition]
\label{ass:local_linear_main}
There is a partition $\{P_1^*, \ldots, P_{K^*}^*\}$ of $\X$, realizable by a finite sequence of axis-aligned binary splits, such that $\tau(x) = \alpha_k^* + x^\top \beta_k^*$ with $\|\beta_k^*\|_0 \le s$ on each $P_k^*$.
\end{assumption}

\begin{assumption}[Minimum subgroup signal]
\label{ass:signal_main}
For distinct regions $k \ne k'$, $\E[(\tau_k(X) - \tau_{k'}(X))^2 \mid X \in P_k^* \cup P_{k'}^*] \ge \Delta_{\min}$ for some $\Delta_{\min} > 0$, where $\tau_k(x) = \alpha_k^* + x^\top \beta_k^*$.
\end{assumption}

\begin{assumption}[Cell regularity]
\label{ass:cells_main}
The number of cells $J = J_n$ satisfies $J \to \infty$ and $n/J \to \infty$.
Each cell has diameter $O(J^{-1/p})$, contains between $c_1 n/J$ and $c_2 n/J$ observations, and has a within-cell covariance whose smallest eigenvalue is at least $c_0\,\mathrm{diam}(C_j)^2$.
The number of cells within $a$ diameters of the true partition boundaries is $O(a J^{1-1/p})$.
\end{assumption}

Condition \ref{ass:bounded_main} is standard in the analysis of honest trees.
Condition \ref{ass:teacher_main} is deliberately teacher-agnostic.
Causal forests, T-learners, and X-learners satisfy it with rates that depend on the smoothness of the effect and of the response surfaces, and the results depend on the teacher only through $r_n$.
Condition \ref{ass:local_linear_main} is the working model of Section~\ref{sec:setup}, with $\beta_k^* = 0$ recovering a piecewise-constant effect.
Condition \ref{ass:signal_main} rules out regions whose effects coincide, which no partition criterion could separate.
Condition \ref{ass:cells_main} asks the cells to be fine enough to resolve the partition boundaries while retaining enough data and enough covariate spread for stable local fits, and its boundary-count clause holds for regular cell coverings because the boundaries are unions of axis-aligned faces.

\subsection{Honest Inference Survives Co-supervision}

The first result shows that the honest leaf estimates remain valid when the tree is built with co-supervision.

\begin{theorem}[Honest leaf-wise ATE inference]
\label{thm:honest_main}
Suppose \ref{ass:randomization} and \ref{ass:bounded_main} hold with a constant propensity score $e(x) \equiv e_0 \in (0,1)$, the tree $\calT$ is constructed from $\mathcal{S}^{\mathrm{con}}$ alone, and $\mathcal{S}^{\mathrm{est}}$ is an independent estimation sample of size $n_{\mathrm{est}}$.
For each leaf $P_k$ with $\Prob(X \in P_k) > 0$, let $\bar\tau_k = \E[\tau(X) \mid X \in P_k]$ and let $\wh\tau_k^{\mathrm{est}}$ be the estimator \eqref{eq:honest_ate}.
Then (a) $\E[\wh\tau_k^{\mathrm{est}} \mid \calT, n_{k,1}, n_{k,0}] = \bar\tau_k$ whenever $n_{k,1} \wedge n_{k,0} \ge 1$; (b) $\sqrt{n_k}\,(\wh\tau_k^{\mathrm{est}} - \bar\tau_k) \to_d N(0, \sigma_k^2)$ as $n_{\mathrm{est}} \to \infty$, where $n_k = n_{k,1} + n_{k,0}$ and $\sigma_k^2 = \Var(Y(1) \mid X \in P_k)/e_0 + \Var(Y(0) \mid X \in P_k)/(1 - e_0)$; and (c) the interval \eqref{eq:ci} has asymptotic coverage $1 - \alpha$ whenever $\sigma_k^2 > 0$.
\end{theorem}

The proof is given in Appendix~B of the Supporting Information.
All of the co-supervision happens on the construction sample, which is where the teacher is trained, the synthetic data are drawn, and the disparity criterion and the threshold refinement are evaluated.
The estimation sample enters only through the leaf averages in \eqref{eq:honest_ate}.
Conditional on the construction sample the tree is therefore a fixed partition, so the honest-estimation argument of Athey and Imbens applies to it unchanged, and nothing in the proof depends on how the partition was built.
The constant-propensity condition makes the treated subjects within a leaf a random sample of the leaf population.
With a variable propensity an inverse-probability-weighted estimator would replace \eqref{eq:honest_ate}.

\subsection{When Co-supervision Helps}

The second result concerns a single cell and asks how the error of its co-supervised model depends on the two data sources.
It is stated for the cell fit of Section~\ref{sec:local_cate} with the Lasso penalty switched off, with the teacher, the pilot, and the cell partition held fixed, and under two structural conditions given in Appendix~A of the Supporting Information: a cross-fitting condition, that the teacher, the pilot, and the partition are fitted on data that do not enter the cell fit, and a moment-matching condition, that the synthetic sampler reproduces the within-cell second moments.

\begin{theorem}[Co-supervision bias-variance decomposition]
\label{thm:cosup_main}
Suppose \ref{ass:randomization} to \ref{ass:cells_main} and the two structural conditions hold.
For a cell $C_j$ with $n_j$ observed and $n_{\mathrm{syn}}$ synthetic points that is contained in a true region and separated from its boundary by at least $\mathrm{diam}(C_j)\sqrt{2\log n}$, the in-cell mean squared error of the unpenalized co-supervised fit $\wh g_j^{\mathrm{co}}$ with weights $(w_{\mathrm{obs}}, w_{\mathrm{syn}})$ satisfies
\begin{equation}
\label{eq:cosup_main}
    \mathrm{MSE}_j(\wh g_j^{\mathrm{co}}) = \frac{w_{\mathrm{obs}}^2\, V_{\mathrm{obs},j}}{n_j} + \frac{w_{\mathrm{syn}}^2\, V_{\mathrm{syn},j}}{n_{\mathrm{syn}}} + w_{\mathrm{syn}}^2\, B_{\mathrm{proj},j}^2 + R_{n,j},
    \qquad R_{n,j} = o_p(n_j^{-1} + r_n^2),
\end{equation}
where $V_{\mathrm{obs},j}$ is $p + 1$ times the average conditional variance of the pseudo-outcome in the cell, $V_{\mathrm{syn},j}$ is the variability of the teacher's error over the synthetic draws, and $B_{\mathrm{proj},j}^2$ is the projection of the teacher's local bias onto the linear model class, bounded by $B_{\mathrm{teacher},j}^2 = \E_{Q_j}[(\wh\tau^{\mathrm{BB}}(x) - \tau(x))^2] = O_p(r_n^2)$, where $Q_j$ denotes the synthetic sampling distribution.
For cells that straddle a region boundary the same expression holds as an upper bound up to a constant factor, with an additive within-cell approximation error $B_{\mathrm{approx},j}^2 = O(L^2 J^{-2/p})$ that also enters the observed-only fit.
The observed-only fit ($w_{\mathrm{syn}} = 0$) has $\mathrm{MSE}_j = V_{\mathrm{obs},j}/n_j + B_{\mathrm{approx},j}^2 + o(n_j^{-1})$.
\end{theorem}

\begin{corollary}[When co-supervision helps]
\label{cor:helps_main}
Up to the remainder terms, co-supervision lowers the cell-level MSE whenever
\begin{equation}
\label{eq:helps_main}
    w_{\mathrm{syn}}^2 \Big( B_{\mathrm{teacher},j}^2 + \frac{V_{\mathrm{syn},j}}{n_{\mathrm{syn}}} \Big) < (1 - w_{\mathrm{obs}}^2)\, \frac{V_{\mathrm{obs},j}}{n_j}.
\end{equation}
When the teacher is consistent with $n_j r_n^2 \to \infty$ and $n_{\mathrm{syn}}$ is large enough that $V_{\mathrm{syn},j}/n_{\mathrm{syn}} = o_p(r_n^2)$, the MSE-minimizing weight satisfies $w_{\mathrm{syn}}^* \asymp V_{\mathrm{obs},j}/(n_j r_n^2)$ and the co-supervised fit dominates the observed-only fit for all sufficiently large $n$.
\end{corollary}

Proofs of Theorem~\ref{thm:cosup_main} and Corollary~\ref{cor:helps_main} are given in Appendix~B of the Supporting Information.
The three terms in \eqref{eq:cosup_main} come from three sources.
The first is the sampling noise of the pseudo-outcomes, which the teacher removes in proportion to the weight it receives.
The second and third are what the teacher brings in, the variability of its error over the synthetic draws, which vanishes as $n_{\mathrm{syn}}$ grows, and its local bias, which does not.
Co-supervision therefore lowers the error when the bias and variance the teacher introduces are smaller than the noise it removes, and the optimal weight grows as the teacher improves.
The adaptive weights \eqref{eq:adaptive_weight} are a plug-in proxy for this trade-off because the cell-level discrepancy $D_j$ estimates $V_j + B_{\mathrm{teacher},j}^2$.
A penalized analogue with the same three-source structure holds for the Lasso-fitted leaf models, with the variance terms of order $p$ replaced by terms of order $s \log p$ (Proposition~B.1 in the Supporting Information).

\subsection{The Disparity Criterion Recovers the Partition}

The last result shows that the disparity criterion of Section~\ref{sec:tree}, applied to accurate cell models, separates the regions of the true partition.

\begin{proposition}[Consistency of the disparity criterion]
\label{prop:disparity_main}
Suppose \ref{ass:randomization} to \ref{ass:cells_main} and the cross-fitting condition of Theorem~\ref{thm:cosup_main} hold, the fitted cell models are uniformly bounded, and they are uniformly accurate on the support $S_t$ of each node, in the sense that $\max_j \sup_{x \in S_t} |\wh g_j(x) - \tau_{k(j)}(x)| \le \epsilon_n \to_p 0$ for cells $C_j$ inside a true region $k(j)$.
Let $\zeta_n \to_p 0$ denote the uniform fluctuation of cell-pair averages, whose leading terms are $\sqrt{J \log J / n}$ and $\sqrt{\log J / n_{\mathrm{syn}}}$.
Then (a) for two distinct regions represented in a node, the disparity \eqref{eq:cate_disparity} averaged over cross-region cell pairs is at least $\kappa_0 \Delta_{\min} - 4\epsilon_n^2 - \zeta_n$, where $\kappa_0$ depends only on the cell-size constants of \ref{ass:cells_main}; (b) for two cells in the same region it is at most $4\epsilon_n^2$; and (c) with probability tending to one, the split selected by \eqref{eq:gos} at a node containing two regions leaves unseparated at most a fraction $C(\epsilon_n^2 + J^{-1/p}\sqrt{\log J} + \zeta_n)/\gamma$ of the cross-region cell pairs whose average true contrast exceeds $\gamma$, the goodness of split at such nodes remains positive, and at pure nodes it is at most $4\epsilon_n^2$.
\end{proposition}

The proof is given in Appendix~B of the Supporting Information.
The proposition uses the cell models only through the uniform accuracy level $\epsilon_n$.
Theorem~\ref{thm:cosup_main} controls the prediction error within a cell at rate $J/n + r_n^2$, whereas uniformity over the node support also requires that fitted slopes do not amplify this error when a model is evaluated away from its own cell, which the Lasso shrinkage in \eqref{eq:local_lasso} and the smoothness of the teacher surface deliver in practice.
We state uniform accuracy as an explicit condition rather than deriving it from primitive assumptions.
Part (c) counts only pairs whose average contrast exceeds $\gamma$ because cell pairs whose regional effects nearly coincide cannot be separated by any disparity-based criterion, and misassigning them costs at most $\gamma$.
The $J^{-1/p}$ term is the fraction of cells that straddle region boundaries, which CATE-weighted clustering shrinks by aligning cells with the teacher's effect surface.

The threshold refinement of Section~\ref{sec:refit} does not change this result.
Its candidate thresholds lie strictly between the largest left-child and the smallest right-child cell centroid, so every cell keeps the child that \eqref{eq:gos} assigned it and the partition of cells to which the proposition refers is unchanged.
When the cells on either side of the split are interior to their regions, the true boundary lies in that same gap, so refinement can only move the threshold toward it.

Each of the three results is checked in Section~\ref{sec:simulations}.
The coverage of the honest intervals in Tables~\ref{tab:coverage} and~\ref{tab:coverage_leafsize} tests Theorem~\ref{thm:honest_main}.
The ablation that removes co-supervision, the comparison of adaptive and fixed weights, and the robustness checks with a corrupted teacher and with a varying number of synthetic points in the Supporting Information test Theorem~\ref{thm:cosup_main} and Corollary~\ref{cor:helps_main}, since each of them varies one quantity in \eqref{eq:cosup_main} while holding the others fixed.
The partition recovery in Table~\ref{tab:ari} and the effect of CATE-weighted clustering in Figure~\ref{fig:lambda_effect} test Proposition~\ref{prop:disparity_main}, and the regions the method merges on the four-quadrant design are the low-contrast pairs that part~(c) leaves unseparated.

\section{Simulation Studies}
\label{sec:simulations}

We evaluate CausalSynthTree through six simulation scenarios that test different aspects of the method, including subgroup recovery under tree-structured and non-tree CATEs, inference quality, and robustness to the null.
The main comparisons use 200 independent replications, and the sensitivity and ablation analyses of Section~\ref{sec:sim_sensitivity} and the Supporting Information use 30 to 100 replications because each replication refits the full pipeline over a grid of settings.

\subsection{Experimental Setup}
\label{sec:sim_setup}

\paragraph{Data-generating processes.}
All DGPs share the structure $Y = \mu(X) + T \cdot \tau(X) + \varepsilon$ with $X \sim N(0, I_p)$, $T \sim \mathrm{Bernoulli}(0.5)$, and $\varepsilon \sim N(0, 1)$.
Unless stated otherwise, $p = 10$.
The six DGPs are summarized in Table~\ref{tab:dgps}.

\begin{table}[t]
\centering
\caption{Summary of data-generating processes.}
\label{tab:dgps}
\small
\begin{tabular}{llll}
\toprule
DGP & $\mu(x)$ & $\tau(x)$ & True partition \\
\midrule
1.\ Simple tree & $x_1 + x_2 + x_3$ & $3 \cdot \mathbf{1}(x_1 > 0) + 1$ & 2 regions by $x_1$ \\
2.\ Linear & $2x_3 + x_4^2$ & $1 + 0.5x_1 + 0.8x_2$ & 2 (by median $\tau$) \\
3.\ Complex tree & $\sin(\pi x_3 x_4) + 2x_5$ & quadrant-wise$^*$ & 4 quadrants of $(x_1, x_2)$ \\
4.\ Smooth & $x_1 + x_2 + x_3$ & $\sin(\pi x_1 x_2)$ & quartiles of $\tau$ \\
5.\ Null & $x_1^2 + x_2^2 + x_1 x_3$ & $2$ (constant) & 1 (all) \\
6.\ Tree-linear & $x_1 + x_2 + x_3$ & $3\,\mathrm{sign}(x_1) + 1.5 x_2$ & 2 regions by $x_1$ \\
\bottomrule
\multicolumn{4}{l}{\footnotesize $^*$DGP 3: $\tau = 2$ if $x_1 > 0, x_2 > 0$; $\tau = 1$ if $x_1 > 0, x_2 \le 0$; $\tau = -1$ if $x_1 \le 0, x_2 > 0$; $\tau = 0$ otherwise.}
\end{tabular}
\end{table}

DGPs 1 and 3 have piecewise-constant tree-structured CATEs, matching CausalSynthTree's assumption~\ref{ass:local_linear_main} with $\beta_k^* = 0$.
DGP~2 has a globally linear CATE with no tree partition, testing whether CausalSynthTree correctly produces a single leaf with a sparse linear model.
DGP~4 has a smooth, non-tree CATE (inspired by Setup~A of \citealt{Nie2021}), deliberately misspecifying all tree-based methods.
DGP~5 has a constant treatment effect (null heterogeneity) with a complex prognostic function $\mu(x)$, testing whether methods avoid confusing prognostic with predictive heterogeneity.
DGP~6 combines a threshold split on $x_1$ with a within-region linear drift, $\tau(x) = 3\,\mathrm{sign}(x_1) + 1.5\,x_2$, the within-region effect modification our sparse-linear leaves target and the case a piecewise-constant tree cannot represent.
This pattern is common in clinical settings, where an effect jumps at a cutoff and then drifts with a biomarker.

\paragraph{Competing methods.}
We compare CausalSynthTree with eight methods spanning the main families in the literature (Table~\ref{tab:methods}). Among these, the causal distillation tree (CDT), run through the \texttt{causalDT} package, distills the same causal-forest teacher CausalSynthTree uses into a constant-leaf CART, so the comparison isolates what the co-supervised sparse-linear leaves add over a plain distillation of that teacher, with the CART-pseudo baseline playing the role of a generic post-hoc extractor.
The causal tree (CT) of Athey and Imbens, run through the \texttt{causalTree} package with the CT splitting rule, the honest splitting criterion, honest cross-validation, and 1-SE pruning, is the interpretable method closest to CausalSynthTree in its inferential aims, since it too pairs a tree with honest leaf estimates.
Its leaf effects for the accuracy metrics are taken from the construction sample, as for every other partition method, and its confidence intervals from the estimation sample.
The modified covariate method fits a single global linear CATE model with no partition, so it cannot represent a tree-structured effect, and its error on the tree-structured DGPs is many times that of CausalSynthTree at every sample size.
We therefore omit it from the simulation tables and retain it in the trial comparison of Section~\ref{sec:realdata}, where a global linear CATE model is the natural benchmark for a single-leaf output.

\begin{table}[t]
\centering
\caption{Competing methods. Random forests use \texttt{scikit-learn} \citep{Pedregosa2011}, causal forests use \texttt{econml} \citep{econml}, and CDT uses the authors' \texttt{causalDT} package.}
\label{tab:methods}
\small
\begin{tabular}{llll}
\toprule
Method & Reference & Implementation & Partition? \\
\midrule
CausalSynthTree & This paper & \texttt{causal\_synthtree.py} & Yes \\
Causal Forest & \citet{Wager2018} & \texttt{econml} & No \\
T-learner & \citet{Kuenzel2019} & \texttt{scikit-learn} RF & No \\
X-learner & \citet{Kuenzel2019} & \texttt{scikit-learn} RF & No \\
Virtual Twins & \citet{Foster2011} & RF $\to$ CART & Yes \\
Mod.\ Covariates & \citet{Tian2014} & Lasso & No (global) \\
CART-pseudo & --- & CART on $Z_i$ & Yes \\
CDT & \citet{Huang2025} & \texttt{causalDT} & Yes \\
Causal Tree & \citet{Athey2016} & \texttt{causalTree} & Yes \\
\bottomrule
\end{tabular}
\end{table}

For methods that do not produce a partition (causal forest, T-learner, X-learner), we obtain subgroup assignments by $k$-means clustering of the estimated $\wh\tau(x)$ for the purpose of computing the adjusted Rand index.
To ensure fair comparison, all methods receive the same honest split: 80\% of the data for model construction, 20\% for estimation-sample inference.
Leaf-wise ATEs and 95\% confidence intervals are computed on the estimation sample using the difference-in-means estimator \eqref{eq:honest_ate}.
Each experiment is repeated 200 times at sample sizes $n \in \{500, 1000, 2000\}$.

\paragraph{Evaluation metrics.}
We report five metrics:
\begin{enumerate}[nosep]
    \item \textbf{CATE MSE:} $n^{-1}\sum_{i=1}^{n}(\wh\tau(x_i) - \tau(x_i))^2$, evaluated on the full dataset.
    \item \textbf{Adjusted Rand Index (ARI):} agreement between the estimated and true partitions, adjusted for chance \citep{Hubert1985}.
    \item \textbf{CI Coverage:} proportion of leaves in which the true leaf-wise ATE falls within the 95\% confidence interval.
    \item \textbf{Best-subgroup bias:} $\wh\tau_{\wh k^*}^{\mathrm{est}} - \bar\tau_{\wh k^*}$, where $\wh k^* = \argmax_k \wh\tau_k^{\mathrm{est}}$.
    \item \textbf{Number of leaves:} tree complexity (for partition-producing methods).
\end{enumerate}

\subsection{Results}
\label{sec:sim_results}

Tables~\ref{tab:cate_mse} to~\ref{tab:coverage} report the main results at $n = 1000$, averaged over 200 replications.
Results at $n \in \{500, 2000\}$, together with the number of leaves of every partition-producing method, are provided in Appendix~D of the Supporting Information.

\subsubsection{Estimation accuracy}

Table~\ref{tab:cate_mse} reports the CATE mean squared error.
Accuracy tracks how closely the leaf model matches the truth.
Where the effect is linear, globally on DGP~2 or within regions on DGP~6, CausalSynthTree is the most accurate method in the comparison, black-box learners included, because the sparse leaf models match the true low-dimensional structure, and the same holds on the coarse two-region DGP~1.
The margin over the constant-leaf distillation is largest on DGP~6, since CDT must keep splitting to approximate a slope it cannot represent, and CausalSynthTree cuts its error by at least two thirds at every sample size with roughly half as many leaves (Table~S6 in the Supporting Information).
The same ordering against CDT holds on two variants of DGP~6 with region-specific modifiers and with four regions (Table~S7 in the Supporting Information), although the X-learner has the lowest pointwise error on both.
Where the truth is a finer partition than the method will grow, on the four-quadrant DGP~3, Virtual Twins, the X-learner, and the causal forest are more accurate pointwise, because CausalSynthTree merges regions with similar effects and returns fewer leaves than the truth requires.
Where the effect is smooth and neither tree-structured nor linear, on DGP~4, it is within a few percent of the causal forest while using fewer than two leaves on average.
The causal tree is the least accurate partition method after CART on pseudo-outcomes, with three to six times the error of CausalSynthTree on every DGP, because it estimates constant leaf effects from the noisy outcomes alone, without a teacher or a within-leaf model.
The null DGP~5 is the setting in which accuracy alone says least, since the tree collapses in most replications and only CDT attains a lower error, by predicting a near-constant effect through many spurious splits, so a low error there must be read alongside the partition recovery in Table~\ref{tab:ari}.

\begin{table}[t]
\centering
\caption{CATE MSE at $n = 1000$ (mean $\pm$ SD over 200 replications). Lower is better. The best result overall is in \textbf{bold}.}
\label{tab:cate_mse}
\small
\setlength{\tabcolsep}{3pt}
\begin{tabular}{l *{8}{C{1.5cm}}}
\toprule
 & CST & CF & X-lrn & VT & T-lrn & \mbox{CART-$Z$} & CDT & CT \\
\midrule
DGP 1 & \textbf{0.064} & 0.079 & 0.157 & 0.135 & 0.422 & 1.739 & 0.102 & 0.220 \\
       & {\tiny (0.066)} & {\tiny (0.035)} & {\tiny (0.033)} & {\tiny (0.048)} & {\tiny (0.063)} & {\tiny (0.456)} & {\tiny (0.054)} & {\tiny (0.217)} \\[2pt]
DGP 2 & \textbf{0.165} & 0.279 & 0.272 & 0.393 & 0.769 & 2.352 & 0.412 & 1.433 \\
       & {\tiny (0.059)} & {\tiny (0.054)} & {\tiny (0.054)} & {\tiny (0.115)} & {\tiny (0.131)} & {\tiny (0.662)} & {\tiny (0.067)} & {\tiny (0.634)} \\[2pt]
DGP 3 & 0.238 & 0.206 & 0.200 & \textbf{0.167} & 0.523 & 1.436 & 0.309 & 0.921 \\
       & {\tiny (0.120)} & {\tiny (0.062)} & {\tiny (0.041)} & {\tiny (0.066)} & {\tiny (0.070)} & {\tiny (0.349)} & {\tiny (0.082)} & {\tiny (0.544)} \\[2pt]
DGP 4 & 0.433 & \textbf{0.416} & 0.427 & 0.472 & 0.576 & 1.183 & 0.432 & 1.084 \\
       & {\tiny (0.016)} & {\tiny (0.017)} & {\tiny (0.031)} & {\tiny (0.061)} & {\tiny (0.060)} & {\tiny (0.212)} & {\tiny (0.019)} & {\tiny (0.591)} \\[2pt]
DGP 5 & 0.032 & 0.047 & 0.255 & 0.183 & 0.694 & 2.949 & \textbf{0.028} & 0.355 \\
       & {\tiny (0.026)} & {\tiny (0.024)} & {\tiny (0.081)} & {\tiny (0.083)} & {\tiny (0.147)} & {\tiny (0.858)} & {\tiny (0.029)} & {\tiny (0.442)} \\[2pt]
DGP 6 & \textbf{0.223} & 0.438 & 0.288 & 0.444 & 0.664 & 1.521 & 1.222 & 1.091 \\
       & {\tiny (0.108)} & {\tiny (0.103)} & {\tiny (0.062)} & {\tiny (0.109)} & {\tiny (0.096)} & {\tiny (0.404)} & {\tiny (0.177)} & {\tiny (0.288)} \\[2pt]
\bottomrule
\multicolumn{9}{l}{\footnotesize CST = CausalSynthTree; CF = Causal Forest; X-lrn = X-learner; VT = Virtual Twins;} \\
\multicolumn{9}{l}{\footnotesize T-lrn = T-learner; CART-$Z$ = CART on pseudo-outcomes; CDT = causal distillation tree;} \\
\multicolumn{9}{l}{\footnotesize CT = causal tree.}
\end{tabular}
\end{table}

\begin{figure}[t]
\centering
\includegraphics[width=\textwidth]{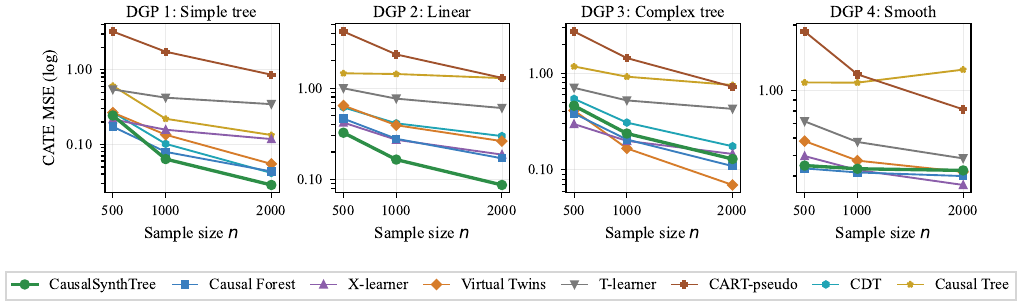}
\caption{CATE MSE versus sample size for DGPs 1 to 4 and every method (log scale, 200 replications). The null DGP~5 and the tree-with-linear-leaves DGP~6 appear in Tables~S1 and~S2 and Figure~S1 of the Supporting Information.}
\label{fig:mse_vs_n}
\end{figure}

Figure~\ref{fig:mse_vs_n} shows the CATE MSE as a function of sample size for DGPs~1 to~4 and every method.
Every method improves with $n$, and CausalSynthTree keeps its lead on the linear DGP~2 at all three sample sizes.
On the simple tree DGP~1 and the tree-with-linear-leaves DGP~6 it trails the best black-box learner at the smallest sample size and overtakes it from $n = 1000$ onward, because its remaining error is the placement of the boundary, which the refined threshold resolves faster with $n$ than a forest resolves a jump.
Its partition recovery improves with $n$ on the tree-structured DGPs, and its coverage remains near nominal at every sample size (Tables~S1 and~S2 in the Supporting Information).

\subsubsection{Partition recovery and interpretability}

Table~\ref{tab:ari} reports the adjusted Rand index between the estimated and true partitions.
Two of the six settings have no tree-structured truth, and there the reference partition is an artificial split of the effect distribution, so a low index records the absence of a partition rather than a failure to find one.
CausalSynthTree scores lowest of all methods on the linear DGP~2 for exactly that reason, because it returns the single leaf the data-generating process calls for.
Among the settings that do have a partition to recover, the method is strongest where the regions are few and well separated.
It is the only method that recovers the single-region truth under the null, where every competing partition method splits into spurious subgroups, the causal tree in more than half of replications and the others in nearly all, and it recovers the two-region partition of DGP~6 best, while every constant-leaf method fragments into extra leaves to chase the within-region slope, a gap that widens with sample size (Table~S6 and Figure~S1 in the Supporting Information).
It is close behind CDT on the simple tree DGP~1 and well ahead of the causal tree and the black-box learners.
Recovery is only partial on the four-quadrant DGP~3, because the method merges regions with similar effects, the low-contrast pairs that part~(c) of Proposition~\ref{prop:disparity_main} allows to stay together, and Virtual Twins ranks higher there by growing a larger tree.

\begin{table}[t]
\centering
\caption{Adjusted Rand index at $n = 1000$ (mean over 200 replications). Higher is better. For CF, T-lrn, and X-lrn, partitions are obtained by $k$-means clustering of $\wh\tau(x)$. The CDT column is run on the identical honest splits with the same causal-forest teacher.}
\label{tab:ari}
\small
\setlength{\tabcolsep}{3pt}
\begin{tabular}{l *{8}{C{1.5cm}}}
\toprule
 & CST & CF & X-lrn & VT & T-lrn & \mbox{CART-$Z$} & CDT & CT \\
\midrule
DGP 1 & 0.929 & 0.689 & 0.507 & 0.455 & 0.486 & 0.412 & \textbf{0.951} & 0.773 \\
DGP 2 & 0.022 & \textbf{0.296} & 0.253 & 0.193 & 0.126 & 0.131 & 0.215 & 0.144 \\
DGP 3 & 0.466 & 0.605 & 0.540 & \textbf{0.702} & 0.266 & 0.393 & 0.576 & 0.396 \\
DGP 4 & 0.002 & 0.008 & 0.024 & 0.031 & \textbf{0.039} & 0.002 & 0.004 & 0.002 \\
DGP 5 & \textbf{0.860} & 0.000 & 0.000 & 0.000 & 0.000 & 0.000 & 0.000 & 0.445 \\
DGP 6 & \textbf{0.954} & 0.524 & 0.498 & 0.287 & 0.472 & 0.350 & 0.505 & 0.420 \\
\bottomrule
\end{tabular}
\end{table}

CausalSynthTree produces the smallest tree of any partition method in every setting but one (Table~S3 in the Supporting Information), the exception being the simple tree DGP~1 at $n = 2000$, where CDT settles at the two true leaves while CausalSynthTree averages slightly more.
On the linear DGP~2 it returns a single leaf in most replications, correctly reporting that the effect is globally linear rather than tree-structured.
Under the null DGP~5 it returns at most two leaves in more than ninety percent of replications at every sample size, while CART on pseudo-outcomes never returns fewer than three, CDT drops below three only at the smallest sample size and in about one replication in eight, the causal tree returns a single leaf in about two fifths of replications but otherwise grows with the sample size, and Virtual Twins drops below three only at the largest.
The second interpretability advantage is the sparse linear model within each leaf.
On DGP~2 the leaf models retain the two true modifiers and give them the two largest coefficients in every single-leaf tree, which is the outcome in nine of ten replications, while the null covariates enter with negligible magnitude (Table~S4 in the Supporting Information).
Methods that produce only pointwise $\wh\tau(x)$ require additional post-hoc tools to communicate which variables drive heterogeneity.

\subsubsection{Inference quality}

Table~\ref{tab:coverage} reports the empirical coverage of 95\% confidence intervals for leaf-wise ATEs on the honest estimation sample.
All methods are within a few points of the nominal level on every DGP.
CausalSynthTree sits at the nominal level throughout, falling marginally above it as often as below, which is what Theorem~\ref{thm:honest_main} predicts, while CART on pseudo-outcomes and the causal tree have the lowest coverage.

\begin{table}[t]
\centering
\caption{Empirical coverage of 95\% CIs for leaf-wise ATEs at $n = 1000$ (mean over 200 replications). Nominal level is 0.950.}
\label{tab:coverage}
\small
\setlength{\tabcolsep}{3pt}
\begin{tabular}{l *{8}{C{1.5cm}}}
\toprule
 & CST & CF & X-lrn & VT & T-lrn & \mbox{CART-$Z$} & CDT & CT \\
\midrule
DGP 1 & 0.943 & 0.936 & 0.940 & 0.932 & 0.950 & 0.923 & 0.935 & 0.933 \\
DGP 2 & 0.965 & 0.943 & 0.943 & 0.944 & 0.947 & 0.921 & 0.932 & 0.939 \\
DGP 3 & 0.939 & 0.936 & 0.943 & 0.935 & 0.945 & 0.912 & 0.938 & 0.925 \\
DGP 4 & 0.963 & 0.943 & 0.944 & 0.939 & 0.948 & 0.928 & 0.930 & 0.935 \\
DGP 5 & 0.955 & 0.932 & 0.937 & 0.926 & 0.931 & 0.924 & 0.929 & 0.941 \\
DGP 6 & 0.943 & 0.931 & 0.929 & 0.922 & 0.931 & 0.917 & 0.925 & 0.908 \\
\bottomrule
\end{tabular}
\end{table}

Table~\ref{tab:coverage_leafsize} stratifies the coverage of CausalSynthTree by the number of estimation-sample observations per leaf.
Coverage is at nominal in leaves with a hundred or more observations.
The smaller bins pool few leaves and scatter on both sides of the nominal level, with the largest shortfall in the smallest leaves of DGP~1, so the normal approximation in Theorem~\ref{thm:honest_main} is reliable at the leaf sizes the method typically produces and should be read with caution below thirty observations.

\begin{table}[t]
\centering
\caption{CausalSynthTree coverage of 95\% CIs stratified by estimation-sample leaf size ($n = 1000$, 200 replications). Cells pool all leaves of the given size across replications, with the number of pooled leaves in parentheses, so cells are unevenly filled. On DGPs~2 and~5 most trees have a single leaf, which falls in the largest bin.}
\label{tab:coverage_leafsize}
\small
\begin{tabular}{l *{6}{C{1.75cm}}}
\toprule
Leaf size & DGP 1 & DGP 2 & DGP 3 & DGP 4 & DGP 5 & DGP 6 \\
\midrule
$<30$ & 0.875 {\tiny (24)} & 1.000 {\tiny (9)} & 0.964 {\tiny (28)} & 1.000 {\tiny (16)} & 1.000 {\tiny (6)} & 1.000 {\tiny (14)} \\
$30$--$60$ & 0.907 {\tiny (54)} & 1.000 {\tiny (10)} & 0.946 {\tiny (112)} & 0.900 {\tiny (30)} & 1.000 {\tiny (10)} & 0.936 {\tiny (47)} \\
$60$--$100$ & 0.939 {\tiny (197)} & 1.000 {\tiny (18)} & 0.905 {\tiny (189)} & 0.953 {\tiny (43)} & 1.000 {\tiny (17)} & 0.939 {\tiny (198)} \\
$100+$ & 0.955 {\tiny (179)} & 0.959 {\tiny (196)} & 0.968 {\tiny (157)} & 0.968 {\tiny (189)} & 0.955 {\tiny (198)} & 0.945 {\tiny (182)} \\
\bottomrule
\end{tabular}
\end{table}

Table~S5 in the Supporting Information compares the naive and de-biased confidence intervals for the best-subgroup ATE on DGPs~1 and~3.
With only two to four candidate leaves the winner's curse is mild, the naive interval still covers near the nominal level, and the de-biased interval, which corrects a bias that is small here, falls a few points below nominal.
The de-biased bootstrap matters more when the number of candidate subgroups is larger.

\subsection{Sensitivity Analysis}
\label{sec:sim_sensitivity}

We examine sensitivity to three design choices, the CATE-weighted clustering parameter $\rho$, the teacher model, and the co-supervision weights.
The three analyses are averaged over 50 to 100 replications at $n = 1000$.

\begin{figure}[t]
\centering
\includegraphics[width=\textwidth]{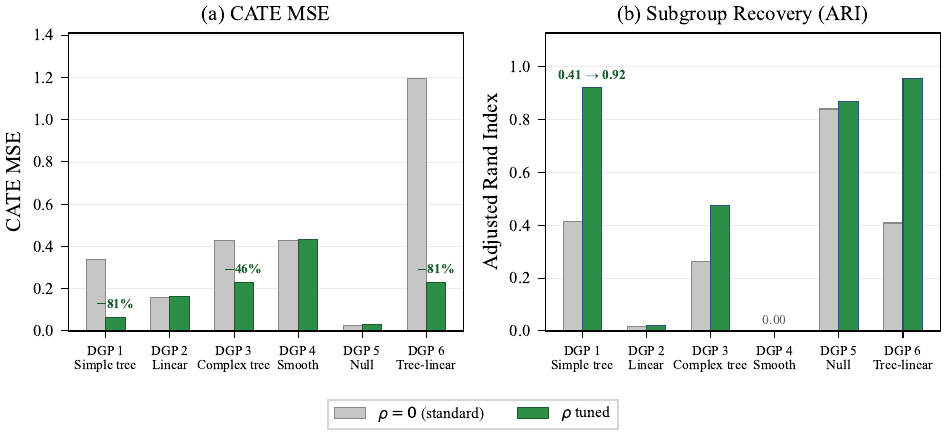}
\caption{Effect of CATE-weighted clustering ($\rho$) on all six DGPs at $n = 1000$ over 100 replications. Percentages are the relative reduction in MSE from $\rho = 0$ to tuned $\rho$, shown where it exceeds 15\%.
Panel~(a): CATE MSE with $\rho = 0$ (standard $k$-means, gray) versus validation-tuned $\rho$ (green).
Panel~(b): adjusted Rand index, same coloring. On the smooth DGP~4 both ARIs are zero, so no bars appear.}
\label{fig:lambda_effect}
\end{figure}

Figure~\ref{fig:lambda_effect} and Table~S9 in the Supporting Information compare CausalSynthTree with standard covariate-based clustering ($\rho = 0$) against validation-selected $\rho \in \{0, 1, 3\}$.
CATE-weighted clustering cuts the error by four fifths on the simple tree DGP~1 and on the tree-with-linear-leaves DGP~6 and by nearly half on the four-quadrant DGP~3, and it roughly doubles the partition recovery on all three, while on the linear, smooth, and null DGPs the validation step selects a value that leaves performance essentially unchanged.

Table~S10 in the Supporting Information compares causal forest and T-learner as the co-supervising teacher.
The causal forest yields lower error on all five DGPs in that table, with the largest gaps on the four-quadrant and null DGPs, where the noisier T-learner manufactures spurious heterogeneity, and the two are comparable on the linear DGP~2.
The tree-with-linear-leaves DGP~6 is the exception, and there the T-learner is the better teacher (Table~S12 in the Supporting Information), because its separate response surfaces track the within-region slope more faithfully than the forest.

Table~S11 in the Supporting Information compares the adaptive per-cell weights \eqref{eq:adaptive_weight}, the plug-in proxy for the optimal weight of Corollary~\ref{cor:helps_main}, with two fixed settings.
The three settings agree to within about ten percent on every DGP except the linear DGP~2, where the even split is about a fifth better than the adaptive weights because the unbiased pseudo-outcomes carry a clean linear signal that deserves more weight than the teacher.
The fixed $(0.3, 0.7)$ setting, which is also the fallback the adaptive rule uses in cells with too few points, tracks the adaptive weights closely, so the weight rule is not a tuning burden once the teacher is accurate.

The three choices above tune the pipeline.
A separate question is how much each structural component contributes, which a unified ablation in Appendix~F of the Supporting Information answers by removing or replacing one component at a time on all six DGPs.
CATE-weighted clustering is what recovers the partition, and the sparse-linear leaf models together with the co-supervision that feeds them are what deliver the accuracy wherever there is structure within regions, since the noisy pseudo-outcomes alone cannot pin down the within-region slopes.
These components are intrinsic to the method rather than additional tuning parameters.
The same appendix reports four further checks.
The default clip rule for the co-supervision weight matches or beats both the bias-corrected rule of Section~\ref{sec:local_cate} and a plug-in oracle weight.
Results are insensitive to the $\rho$ grid, and a fixed $\rho = 2$ is a reliable default when the validation slice is small.
Root-level splits are stable, and performance is flat over a wide range of the number of synthetic points.

\section{Real Data Applications}
\label{sec:realdata}

We apply CausalSynthTree to two randomized trials that differ in exactly the way that matters for subgroup analysis.
In the first, no covariate shows any sign of effect modification, so the right answer is to report none.
In the second, one covariate carries a borderline signal, so the right answer is to report it and say how weak it is.
In both trials we judge every method against the same external yardstick, a pre-specified test that adds a single treatment-by-covariate interaction to a covariate-adjusted linear model and therefore involves no selection.
Those tests are collected in Appendix~G of the Supporting Information.

\subsection{The ACTG175 HIV Trial}
\label{sec:actg}

\subsubsection{Trial data and pre-specified tests}
The ACTG175 trial \citep{Hammer1996} enrolled $n = 2139$ patients randomized to four arms, zidovudine monotherapy ($n = 532$), zidovudine with didanosine ($n = 522$), zidovudine with zalcitabine ($n = 524$), and didanosine monotherapy ($n = 561$).
Following \citet{Foster2011} and \citet{Athey2016} we compare monotherapy ($T = 0$, $n = 532$) with combination therapy ($T = 1$, $n = 1607$), which inherits a 3 to 1 randomization ratio, so the pseudo-outcomes and validation losses use the design propensity $e(x) = 3/4$ throughout.
The outcome is the change in CD4 count from baseline to 20~weeks, and we use the 12 baseline covariates listed in Appendix~G of the Supporting Information.
The naive average treatment effect is $50.4$ CD4 cells/mm$^3$.
We apply honest estimation with an 80/20 construction and estimation split.

The pre-specified tests find nothing.
A joint test of all twelve treatment-by-covariate interactions gives $\chi^2 = 12.1$ on 12 degrees of freedom, $p = 0.44$, a statistic almost exactly equal to its degrees of freedom.
No single covariate comes close, the smallest being $p = 0.11$, and the trial's own randomization stratification factors, which are not among the 12 model covariates, are tested separately and give $p = 0.93$ for prior antiretroviral experience and $p = 0.80$ for recent zidovudine use.
The largest contrast anyone would look for, between patients above and below a baseline CD4 count of 300, is about five CD4 cells/mm$^3$ against an overall effect of about fifty.

\subsubsection{CausalSynthTree output and power calibration}
CausalSynthTree returns a single-leaf tree in all 50 random seeds, with $\rho$ and $J$ selected by validation $R$-loss.
Every candidate split is removed by pruning, so the method reports no tree-structured subgroups.
The output is a single sparse linear CATE model for the whole population, and that model is nearly constant.
In the reference seed the Lasso retains only months of prior antiretroviral treatment, with a coefficient of $+0.01$, and sets the remaining 11 coefficients to zero, and across seeds no covariate is retained often enough, or with a large enough coefficient, to report.
The honest estimate of the overall effect on the estimation sample is $39.7$ CD4 cells/mm$^3$ with a 95\% confidence interval of $[15.7,\; 63.8]$, which differs from the naive full-sample figure only by sampling variation, since the interval covers it.
Across the 50 seeds the honest leaf effect has mean $48.0$ with standard deviation $11.2$, and it is positive in every seed.
The finding does not depend on the size of the construction sample, since halving it to a 50/50 split still gives a single leaf in 48 of 50 seeds (Appendix~G of the Supporting Information).
Set against the pre-specified tests, the single leaf is the correct answer, not a conservative one.

A plasmode power check \citep{Franklin2014} on the real covariates (Appendix~H of the Supporting Information) calibrates the null finding.
With outcomes simulated at the trial's noise level and a known CD4-threshold subgroup injected, the same pipeline holds its false-split rate at 2\%.
It begins to detect the subgroup only once the contrast reaches twice the trial's overall average effect, and finds it in about half of replications at three times.
At those contrasts every replication that splits recovers the correct variable and a threshold close to the truth, while CART on pseudo-outcomes splits in every replication even when no subgroup exists.
The single-leaf result is therefore evidence against a subgroup contrast of that size, and smaller contrasts are below what this trial's noise level allows any procedure with controlled false-positive rate to detect.

\subsubsection{Competing methods}
The competing partition methods disagree, confidently and with each other.
Over the same 50 splits, Virtual Twins never returns fewer than four leaves and roots its tree on baseline CD4 in 34 seeds, while CART on pseudo-outcomes never returns fewer than four leaves and roots on months of prior antiretroviral treatment in all 50.
Each of these two is internally reproducible, so instability is not the problem.
The problem is that they name different modifiers, and the direct tests rule out both, with $p = 0.67$ for baseline CD4 and $p = 0.27$ for months of prior therapy.
This is what a stable false discovery looks like.
The causal distillation tree, distilling the same causal-forest teacher CausalSynthTree uses, fails in a third way.
It never returns fewer than five leaves, about eleven on average, and its root moves among baseline CD4, weight, and months of prior therapy from split to split, so it asserts subgroup structure in every split without agreeing on which covariate defines it.
Both failure modes motivate the pruning and false-positive control built into CausalSynthTree.
Table~\ref{tab:actg_compare} compares the fitted effects across methods.
All agree on a positive mean effect of similar size, CausalSynthTree sits with the causal forest at the conservative end, and its predicted heterogeneity is the smallest by a wide margin.
The pairwise correlations of their predictions (Table~S20 in the Supporting Information) show moderate agreement within the plug-in family of causal forest, X-learner, and T-learner and weak agreement between every other pair.

\begin{table}[t]
\centering
\caption{Comparison of CATE predictions on the ACTG175 trial.}
\label{tab:actg_compare}
\small
\begin{tabular}{lcccc}
\toprule
Method & Mean $\wh\tau$ & SD $\wh\tau$ & Leaves & Interpretable? \\
\midrule
CausalSynthTree & 48.0 & 2.6 & 1 & Yes (sparse linear) \\
Causal Forest & 47.3 & 15.9 & --- & No \\
X-learner & 52.9 & 26.8 & --- & No \\
Virtual Twins & 51.1 & 15.2 & 5 & Yes (tree) \\
T-learner & 51.0 & 45.0 & --- & No \\
CART-pseudo & 53.2 & 32.9 & 5 & Yes (tree) \\
CDT & 51.4 & 8.5 & 10 & Yes (tree) \\
MC-Lasso & 51.5 & 14.2 & 1 & Yes (linear) \\
\bottomrule
\end{tabular}
\end{table}

\subsection{The Moertel Colon Cancer Trial}
\label{sec:colon}

\subsubsection{Trial data and pre-specified tests}
The second application is the adjuvant therapy trial of \citet{Moertel1990}, distributed with the \texttt{survival} package for R \citep{Therneau2023}.
Patients with resected colon carcinoma were randomized in equal proportions to observation, levamisole alone, or levamisole with fluorouracil.
We compare observation with the levamisole and fluorouracil arm, the combination that the trial established as effective, so that $e(x) = 1/2$ by design.
The outcome is recurrence within five years, a binary endpoint for which only 5\% of patients are censored before the horizon.
Those patients are dropped, leaving $n = 563$ with complete covariates, of whom 274 received the combination.
Nine baseline covariates are used, listed in Appendix~G of the Supporting Information.
The overall effect is large and unambiguous, a five-year recurrence risk of $0.57$ under observation against $0.40$ under treatment.
We again apply honest estimation with an 80/20 split.

Here the pre-specified tests give a different answer from the HIV trial, though a guarded one.
The omnibus test over all nine interactions is not significant, $\chi^2 = 8.2$ on 9 degrees of freedom, $p = 0.52$, but one covariate stands out on its own.
The treatment-by-sex interaction is $-0.156$ with $p = 0.052$, against an unadjusted five-year risk difference of $-0.08$ among women and $-0.28$ among men.
Since this is one of nine covariates examined, the finding does not survive any correction for multiplicity, and the same interaction on five-year mortality is in the same direction but weaker, $p = 0.15$.
A sex difference in the benefit of fluoropyrimidine therapy is biologically discussable rather than absurd, but on this evidence it is a hypothesis, not a result.
The question for a subgroup method is whether its output reflects that strength of evidence, neither manufacturing a definitive subgroup nor discarding the signal entirely.

\subsubsection{CausalSynthTree output}
CausalSynthTree returns a single leaf in 31 of 50 seeds.
In the 19 seeds where it does split, sex is the most frequent root, chosen more often than any other covariate, and the remaining choices are scattered across age, differentiation grade, extent of spread, and nodal status.
The method therefore splits about a third of the time, on the one covariate the direct test flags, which is the graded response the evidence warrants.
Table~\ref{tab:colon_tree} displays the fitted tree from the first such seed, the smallest tree rooted on sex.
Two features of it are worth noting.
The honest intervals separate in the direction the raw data suggest, with a significant reduction in five-year recurrence risk among men and an interval covering zero among women.
And the two leaf models behave differently.
The model for women is nearly constant, while the model for men carries an age gradient spanning about $0.28$ in risk difference across the observed age range, so that younger men are predicted to benefit most.
That gradient is not an artifact of the fitting procedure.
The pre-specified treatment-by-age interaction, estimated without reference to the tree, is $-0.005$ per year of age with $p = 0.17$, and the leaf model recovers a slope of $-0.004$ per year.
A within-region effect modifier of this kind is precisely what a constant-leaf distillation cannot represent, and it is the reason the leaves carry sparse linear models.
The honest estimates rest on 111 estimation-sample patients and are correspondingly noisy, and we present the tree as the object a practitioner would receive, together with the caveat that the evidence behind the split is borderline.

\begin{table}[t]
\centering
\caption{The fitted CausalSynthTree on the colon cancer trial, from the reference seed. The effect is the five-year recurrence risk difference, so negative values favor treatment. Honest estimates use the held-out estimation sample, the mean fitted CATE summarizes the leaf model over the patients routed to the leaf, and the last column is the raw contrast in the full trial for context.}
\label{tab:colon_tree}
\small
\begin{tabular}{llcccc}
\toprule
Leaf & Region & $n_{\mathrm{est}}$ & Honest effect (95\% CI) & Mean fitted CATE & Full-trial contrast \\
\midrule
1 & Women & 60 & $-0.13$ $[-0.39,\; +0.12]$ & $-0.08$ & $-0.08$ \\
2 & Men & 51 & $-0.35$ $[-0.61,\; -0.09]$ & $-0.21$ & $-0.28$ \\
\bottomrule
\multicolumn{6}{l}{\footnotesize The model for women is nearly constant. The model for men declines by about $0.004$ in} \\
\multicolumn{6}{l}{\footnotesize risk difference per year of age, so the predicted benefit is largest in younger men.}
\end{tabular}
\end{table}

\subsubsection{Competing methods}
The competing tree methods fail in different directions on this trial.
CART on pseudo-outcomes splits in every seed and reports about five leaves, but its root variable is unstable across outcomes rather than across seeds, choosing sex in 30 of 50 seeds for recurrence and nodal status in 30 of 50 for mortality, and nodal status shows no interaction at all, with $p = 0.80$ and $p = 0.35$.
The causal distillation tree also splits in every seed, into about six leaves, rooted on sex in about half of the recurrence splits and on age in most of the rest, and on mortality its root moves to age and nodal status.
It therefore lands on the right covariate more often than not, but presents the finding as a six-leaf partition, with no indication that the evidence behind it is borderline.
Virtual Twins goes the other way and returns a single leaf in 42 of 50 seeds, so on this trial it would report no subgroups and miss the one signal present.
CausalSynthTree is the only method of the four whose output moves with the strength of the evidence, declining to split on the trial where nothing is there and splitting part of the time, on the right variable, where something might be.

Neither trial shows heterogeneity that a well-powered omnibus test can detect.
That is the normal state of affairs for a trial sized to estimate a main effect, since detecting an interaction of a given magnitude requires roughly four times the sample size needed for a main effect of the same size \citep{Brookes2004}, and it is the regime in which a subgroup method is judged mainly by what it refrains from reporting and by whether the little it does report survives an independent check.
This is the realistic use case for the method, an exploratory analysis whose output can be handed to a clinician with a defensible statement about how much evidence stands behind it, as the ICH E9(R1) framework asks.

\section{Discussion}
\label{sec:discussion}

This paper introduced CausalSynthTree, a method for interpretable subgroup identification that combines co-supervised local CATE estimation with tree-based region synthesis.
Its output is a shallow tree whose leaves carry sparse linear models for the treatment effect, together with honest confidence intervals for the leaf-wise effects that remain valid despite the synthetic, teacher-labeled data used to build the tree.

The simulations mark out where the method belongs, between the black-box learners and the interpretable trees, with the accuracy of the former where the effect has low-dimensional structure and the output of the latter.
Its advantage over the black-box learners is largest when a partition exists and the effect drifts linearly within its regions, and it holds on a single sharp threshold from a thousand observations onward.
When the effect is linear with no partition it returns a single leaf, and the sparse model in that leaf is the most accurate estimate in the comparison.
The advantage disappears on a finer partition whose neighboring regions carry similar effects, which the method merges.
A practitioner who expects many small regions with sharp thresholds and nothing else may therefore prefer a larger constant-leaf tree.
One who needs within-region effect modification, or who must control false discoveries, is the reader this method serves.
The two trials put those regimes in context: judged against pre-specified tests, CausalSynthTree was the only partition method whose output moved with the evidence on both.

The teacher matters in practice, and not only through its pointwise accuracy.
The causal forest is the better teacher on five of the six DGPs, but the T-learner wins on the one with within-region slopes, so the teacher is best chosen by the kind of heterogeneity expected.
The adaptive weights protect against a biased teacher by measuring its disagreement with the unbiased pseudo-outcomes.
Propagating that weight to the leaf refit, an option we call adaptive leaf weighting, keeps the error bounded even under a corrupted teacher and costs nothing when the teacher is good (Figure~S2 in the Supporting Information), and we recommend enabling it when teacher reliability is uncertain.
The bias-corrected weight of Section~\ref{sec:local_cate} does not improve on the clipped weight in our experiments.

The output answers the regulatory expectations set out in Section~\ref{sec:intro}.
Honest inference delivers valid confidence intervals for subgroup-level effects without pre-specifying the subgroups, and the multi-seed stability analysis of Section~\ref{sec:realdata} offers a check on the reproducibility of a discovered subgroup, a concern raised by \citet{Lipkovich2017}.
The two applications show why that check is needed, since the competing tree methods are internally reproducible across splits and still disagree about which covariate defines the subgroup.
The plasmode design of Appendix~H of the Supporting Information adds a way to calibrate what a null finding means at a given trial's noise level.

The method has four limitations worth stating.
Its accuracy depends on the teacher, and although the adaptive weights and adaptive leaf weighting limit the damage from a biased teacher, they cannot recover signal the teacher has smoothed away.
Honest estimation splits the data into construction and estimation samples, which reduces the effective sample size for both tree building and inference and may be too costly in small trials.
The disparity criterion chooses split variables on cell centroids rather than on individual observations, so the candidate set at deep nodes rests on a few cells, although root-level splits are stable in practice and the refinement step recovers the threshold locations from the observations.
Finally, $k$-means produces approximately spherical cells, and a Gaussian-mixture alternative could suit correlated covariates better.

Several extensions are natural.
Survival outcomes can enter through inverse-probability-of-censoring-weighted pseudo-outcomes.
Multi-arm trials call for simultaneous estimation of several CATE functions with a disparity criterion that aggregates over contrasts.
Observational data require an estimated propensity score, for which the pseudo-outcome \eqref{eq:pseudo_augmented} with a fitted $\wh e(x)$ remains consistent under a correctly specified propensity model, and the doubly robust pseudo-outcome of \citet{Kennedy2023} would protect against misspecification of either nuisance.
Combining several teachers within the co-supervision step is a further direction when no single teacher dominates.


\section*{Funding Information}
The research of Evgenii Kuriabov and Jia Li is supported by the National Science Foundation under Grant CCF-2205004.

\section*{Conflicts of Interest}
The authors declare no conflicts of interest.

\section*{Data Availability Statement}
Both trials analyzed in Section~\ref{sec:realdata} are publicly available, ACTG175 in the R package \texttt{speff2trial} \citep{speff2trial} and the colon cancer trial in the R package \texttt{survival} \citep{Therneau2023}.
Python code implementing CausalSynthTree and reproducing all simulations and analyses will be made publicly available on GitHub upon acceptance.

\section*{Supporting Information}
Additional supporting information can be found online in the Supporting Information section.
Appendices~A to~H contain the formal statements and proofs, the pruning algorithm, additional simulation results, the de-biased Lasso evaluation, ablation and robustness analyses, the interaction tests and method stability for both trials, and the plasmode power check, with Tables~S1 to~S21 and Figures~S1 and~S2.

\bibliographystyle{apalike}

\end{document}